\documentclass[aps,prl,english,twocolumn,notitlepage]{revtex4-1}
\usepackage{mathptmx}
\usepackage{graphicx}
\usepackage[T1]{fontenc}
\usepackage[latin9]{inputenc}
\usepackage{amsthm}
\usepackage{amsmath}
\usepackage{amssymb}
\usepackage{blkarray}
\usepackage{xcolor}
\usepackage{tikz}
\usepackage{verbatim}
\usepackage{lipsum}
\usepackage{dsfont}
\newcommand{\newc}{\newcommand}
\newc{\beq}{\begin{equation}}
\newc{\eeq}{\end{equation}}
\newc{\kt}{\rangle}
\newc{\br}{\langle}
\newc{\beqa}{\begin{eqnarray}}
\newc{\eeqa}{\end{eqnarray}}
\newc{\Tr}{\mbox{Tr}}
\newc{\ep}{\text{ep}}
\newc{\ovl}{\overline}
\newc{\longra}{\longrightarrow}
\newc{\ot}{\otimes}
\newc{\evn}{\mathcal{E}_{vn}}

\newc{\ca}{{\cal A}}
\newc{\cb}{{\cal B}}
\newc{\ta}{{\tilde A}}
\newc{\tb}{{\tilde B}}

\newtheoremstyle{mystyle}
  {}
  {}
  {\normalfont}
  {}
  {\itshape}
  {.}
  { }
  {}
\theoremstyle{mystyle}

\makeatletter
\let\Hy@backout\@gobble
\makeatother

\begin{document}
\title{Entangling power of time-evolution operators in integrable and nonintegrable many-body systems}

 \author{Rajarshi Pal }
 \author{Arul Lakshminarayan}
 \affiliation{Department of Physics, Indian Institute of Technology Madras, Chennai, India 600036.}
 
\begin{abstract}

The entangling power and operator entanglement entropy are state independent measures of entanglement. Their growth and saturation is examined in the time-evolution operator of quantum many-body systems that can range from the integrable to the fully chaotic. An analytically solvable integrable model of the kicked transverse field Ising chain is shown to have ballistic growth of operator von Neumann entanglement entropy and exponentially fast saturation of the linear entropy with time. Surprisingly a fully chaotic model with  longitudinal fields turned on shares the same growth phase, and is consistent with a random matrix model that is also exactly solvable for the linear entropy entanglements. However an examination of the entangling power shows that its largest value is significantly less than the nearly maximal value attained by the nonintegrable one. The importance of long-range spectral correlations, and not just the nearest neighbor spacing, is pointed out in  determing the growth of entanglement in nonintegrable systems. Finally an interesting case that displays some features peculiar to both integrable and nonintegrable systems is briefly discussed. 
\end{abstract}

\maketitle

Effects of nonintegrability in quantum systems, dubbed  quantum chaos or chaology \cite{BerryChaology,Haake,Gutzwiller90}, is now being vigorously studied in many-body systems
with motivations ranging from thermalization and
localization transitions to information scrambling \cite{Rigol2016,Pollman14, Roberts2015,Swingle2016,Prosen17,Luitzotoc17}. 
While single body quantum chaos engenders states behaving like those chosen from a uniform distribution over the Hilbert space \cite{Haake}, in many-body contexts it is fair to say that little is known about detailed statistical properties of stationary or time evolving states. One important window into this is provided by various types of entanglements in the states \cite{Horodecki09,Amico08}. 
For example, in bipartite systems a chaotic evolution rapidly entangles the two subsystems to nearly the maximum possible extent even when they are initially product states \cite{MS99,BA02}.

The most well-studied entanglement is that between two halves of many-body systems such as spin-chains, the block-entropy \cite{Eisert2010}. 
Natural questions of importance is how fast such an entanglement grows for initially unentangled states, the extent it reaches or saturates, and what distinguishes integrable and chaotic systems in 
these contexts \cite{Amico08,Viola06,Nahumhaah17,Najafi18,Prosen07}.
In integrable critical systems described by a conformal field theory and the transverse Ising model it was shown to grow linearly with time $t$ before it saturates to a value dependent on the initial state \cite{JC05}. 
Such a ballistic growth of entanglement was also seen, somewhat surprisingly, in a nonintegrable Ising model with longitudinal fields where energy transport itself is diffusive \cite{HK13}. 

Compared with a nonintegrable case, the integrable one may entangle certain initial states faster and to a larger extent, while others lesser.
Thus it is desirable and interesting to consider entanglement measures that are independent of the initial state. Two approaches present themselves, the first wherein the operator entanglement of propagators in time are studied. 
This was recently adopted in \cite{Luitz2017}, where a many-body Floquet nonintegrable system, and the Heisenberg model with disordered fields were considered. It was  shown by simulations that the time-evolution 
operator entanglement entropy shows a linear, power-law and logarithmic growth respectively for the Floquet system, the Heisenberg model in the weak disorder 
phase and the many-body-localized phase \cite{zanardi2000,zanardi2001}. 

The second approach is based on the notion that it is illuminating to look directly into the ability of time evolution operators to create entangled states starting from {\textit{arbitrary}} product states. As state entanglement is an
important resource in information processing tasks, entangling abilities of unitary gates and entanglement of operators as a dynamical resource \cite{ND03} have been considered. In particular,
bipartite entangling power of an unitary operator has been defined as the average entanglement created when acting on a uniform distribution of product states \cite{zanardi2000,zanardi2001}. This was shown in \cite{zanardi2001} 
to be related to  operator entanglement entropy in a non-trivial way when defined via the linear entropy. 
This has been applied previously for example to quantum transport in light-harvesting complexes \cite{Plen10} and characterization of quantum chaos \cite{DD04,Kestner18}. 

Using both approaches here, we study the entangling power and operator entanglement entropy of the unitary time-evolution operators $U(t)$ for simple spin chains in both integrable and non-integrable regimes. Freed from the specificity of the initial state, we study the rate of growth of these quantities, their eventual saturation values if any, and compare them with a random matrix theory (RMT) model. We find analytically, ballistic growth of operator entanglement and entangling power in a particular case of the integrable transverse Ising model, reflecting the ballistic growth of state entanglement \cite{JC05}. As in the case of states it is found that the operator entanglement of certain integrable models can outstrip that of non-integrable models, calling into question the superior entangling capabilities of chaos \cite{MS99,BA02,LS05,Scott03}. However we show that the entangling power in the same models is higher for the nonintegrable case, underlining the role of entangling power as opposed to operator entanglement.

 The RMT model replaces the blocks between which the entanglement is found by random operators while retaining the interaction and is thus a hybrid one which is seen to be sometimes surprisingly good. While it can be expected to work for nonintegrable spin chains, the ballistic growth implied in some cases also leads to coincidence of the RMT with integrable models in the growth phase, partially resolving the ballistic growth seen in both integrable and nonintegrable cases.
 
RMT models serve as a good foil for many non-integrable cases of the spin chains, as there is a correlation between the commonly used spectral property of the nearest neighbor spacings (NNS) and how well the RMT model works. However we also find that there are non-integrable models with Wigner distribution of the NNS but yet do not follow the RMT model for entangling power just as well. This prompts the study of long-range spectral correlations such as number-variance which shows differences amongst these non-integrable cases. This demonstrates that long-range correlations, that affect short-time behaviours, are important in understanding the growth of entanglement in these systems than simply the NNS alone. 

{\it Measures employed:} Most measures of entanglement strengths of an operator $U$ \cite{ND03} acting in a bipartite space $ {\cal H}_A \otimes {\cal H}_B$, $\mbox{dim}({\cal H}_{A,B})=N$ are based on its Schmidt decomposition,
$U= N \sum_{i=1}^{N^2}\sqrt{\lambda_i} \, A_i \otimes B_i$,  
 with $A_i$ and $B_i$ being orthonormal operators on ${\cal H}_{A,B}$ satisfying, $ \Tr(A_i^{\dagger}A_j)=\Tr(B_i^{\dagger}B_j)=\delta_{ij}$, $\lambda_i \geq 0$. If $U$ is unitary, as will be the cases in this work, then $\sum_i \lambda_i = 1$.
We  will consider operator entanglement entropies defined via both the linear and von Neumann entropies as,
\begin{equation}
\label{eq:Elin}
E_l(U)= 1 - \sum_{i=1}^{N^2} \lambda_i^2 , \;\; E_{vN}(U)= -\sum_{i=1}^{N^2} \lambda_i  \log \lambda_i. 
\end{equation} 
These vanish iff $U$ is a tensor product of two operators.

Following, \cite{zanardi2000}, the entangling power of a unitary operator is defined as,
$\ep(U)= \overline{{\cal E}(|\psi \kt = U|\psi_A\rangle |\psi_B\rangle)}^{|\psi_A\rangle, |\psi_B\rangle }$ 
with ${\cal E}$ being a suitable entanglement measure of {\it states} and the average is taken over all the product states $|\psi_A\rangle$, $|\psi_B\rangle$
distributed uniformly. In this paper, we consider ${\cal E}$ to be either the linear or von Neumann entropy of the 
reduced density matrix $\rho_A=\Tr_{B}(|\psi\kt \br \psi|)$, that is ${\cal E}_L=1-\Tr(\rho^2_A)$ and ${\cal E}_{vN}=-\Tr(\rho_A \log \rho_A)$ respectively. The corresponding entangling powers are denoted by $\ep_{l}(U)$ and $\ep_{vN}(U)$.  
 
While the entangling power has a natural interpretation as the average entanglement that is created by the action of $U$ on arbitrary product states, 
it was shown in \cite{zanardi2000} that $\ep_l(U)$ is intimately connected to the operator linear entanglement entropy $E_l(\cdot)$ as,
\begin{equation}
 \ep_l(U)= \frac{N^2}{(N+1)^2} (E_l(U) + E_l(US) - E_l(S)), 
\end{equation}
where $S$ is the swap operator $S|\psi_A\kt|\psi_B\kt = |\psi_B\kt|\psi_A\kt$.
Additionally  we also find $\ep_{vN}(U)$ for which there is no such simple connection known to $E_{vN}(\cdot)$ and hence resort to finding it 
numerically.

To emphasize the qualitative difference between operator entanglement and entangling power of operators, we briefly recall the ancilla interpretation of the former, wherein we 
imagine that $A$ and $B$ are equipped additionally with $N$ dimensional systems $A'$ and $B'$. Let $AA'$ and $BB'$ be in the  standard maximally entangled state $|\phi^+\kt=\sum_{j=1}^N|jj\kt/\sqrt{N}$, and consider 
the 4-party state $|\Phi_U\kt=(U_{AB} \otimes \mathds{1}_{A'B'})|\phi^+\kt _{AA'}|\phi^+\kt _{BB'}$, where $U_{AB}\equiv U$. Then $E_{l,vN}(U)$ ($E_{l,vN}(US)$) are the linear and von Neumann entropies of the reduced
state $\rho_{AA'}=\Tr_{BB'}(|\Phi_U\kt \br \Phi_U|)$ ($\rho_{AB'}= \Tr_{A'B}(|\Phi_U\kt \br \Phi_U|)$). These reduced states can also be related to reshuffling and partial transpose of $U$, allowing for their direct evaluation
\cite{Bhargavi2017}. The central difference therefore is that operator entanglement can be viewed as the entanglement in one particular 4-party state engendered by the action of a bipartite unitary operator, while the entangling power is the average entanglement in an ensemble of states resulting from its action on all 2-party product states. 

Averages over the Haar measure of unitary operators on $ {\cal H}_A \otimes {\cal H}_B$, are important to compare with the saturation values 
for non-integrable models. Thus we state known results \cite{zanardi2000,KZKus,Luitz2017}: 
\beq
\label{eq:RMTavgs}
\begin{split}
&\ovl{\ep_l}=(N-1)^2/(N^2+1),\; \ovl{E_l}=(N^2-1)/(N^2+1),\\ 
&\ovl{E_{vN}} \approx 2\log{N}-1/(2\ln(2)).
\end{split}
\eeq
 While $\ovl{\ep_{vN}}$ is unknown, it is close to the Haar averaged value for pure states \cite{P93} namely $\log(N)-(1/(2\ln(2))$. These are simply referred to ahead as RMT averages.

{\it The spin models:} We consider the following  Floquet Hamiltonian for a spin chain of $L$ sites: $H(t)= H_0 + V  \sum_{k=-\infty}^{+\infty} \delta(k-t/\tau)$, with 
\begin{equation}
\label{model}
H_0=\sum_{j=1}^{L-1} \sigma_j^z \sigma_{j+1}^z + \sum_{j=1}^L h_i^z \sigma_i^z, \;\text{and}\; V= \sum_{j=1}^L \left(h_j^x\sigma_j^x +  h_j^y \sigma_j^y \right),
\end{equation}
a kicked version of the Ising model with a magnetic field which has both transverse and longitudinal components. The model is integrable \cite{LS05} for purely transverse ($h_i^z=0$) or purely longitudinal ($h_i^x=h_i^y=0$) fields  and  nonintegrable otherwise.    
The state (in $\hbar=1$ units) just after the ($n+1$)th kick is connected to the state just after the $n$-th kick by the unitary Floquet operator:
$|\psi(n+1) \rangle = |\psi(n)=U(\tau,h)|\psi(n)\rangle $, with 
\begin{equation}
\label{eq:spinfloquet}
U(\tau,h) = e^{-iV\tau}e^{-iH_0\tau}.
\end{equation}
The present work considers the bipartite entangling power and operator entanglement entropies of $U^n(\tau,h)$ between the first half and the rest of the spins, the dimension of the single party Hilbert space thus being $N=2^{L/2}$. As $\tau \rightarrow 0$ the kicked model goes over to a continuous time evolution; and the discussions below therefore can be extended to time-independent Hamiltonians via the Suzuki-Trotter decomposition of the propagator.

We consider the following nonintegrable and integrable magnetic field configurations referred to as Set-NI: ($h_i^x=0.9045, h_i^y=0.3457,h_i^z=0.8090$) and Set-I: ($h^x_i=1.0, h^y_i=0, h^z_i=0)$ respectively. The 
nonintegrable configuration is chosen mainly for comparison with literature \cite{Luitz2017,HH13} and not because of any fine-tuning. While we consider various values of $\tau$, the case $\tau=\pi/4$, 
is special as we exactly solve for the operator entanglement entropy and entangling power for the integrable case.  Also, for the chosen nonintegrable     
configuration at this value of $\tau$ the Floquet operator seems ``maximally random" going beyond that captured by NNS distribution discussed  in \cite{HH13}. 

{\it The RMT Model:} The spin Floquet operator in Eq.~(\ref{eq:spinfloquet}) is of the form $U(\tau, h)=(U_A \otimes U_B)\, U_{AB}(\tau)$ where $U_{A,B}$ are local to blocks $A$ and $B$ consisting of spins $1,\cdots, L/2$ 
and $L/2+1, \cdots,L$ respectively and $U_{AB}(\tau)=\exp \{-i\tau \sigma_{L/2}^z \sigma_{L/2+1}^z\}$ is the nonlocal interaction between the blocks. While the local operators, which contain all the information of the fields, do
not contribute to the entangling power of $U$ itself, they play a crucial role for the powers $U^n$ we are interested in \cite{Bhargavi2017}.
The RMT model $U_{RMT}(\tau)$ is a hybrid one wherein we replace $U_{A,B}$ by local random unitary matrices and retain the interaction as is. As $U_{A,B}$ are merely $L/2$ length chains of the original kind, this can be expected to be a 
reasonable model if $U(\tau,h)$ is sufficiently nonintegrable and possesses random matrix properties. 

Such models have been used to study spectral and entanglement transitions in coupled chaotic bipartite systems \cite{Shashi16,Arul16}. 
Random circuit models \cite{Emerson04,Kastoryano17} have been used for many-body systems and to study the growth of initially local operator \cite{HaaH18}, and will be useful in the context of entangling power as well. However as we are considering bipartite entanglements and there are analytical results available if the local block operators are fully random we adopt this caricature.

{\it Operator entanglement and entangling power averages:} While quantities such as $ \br E_l[U^n_{RMT}(\tau)] \kt_{loc}$, where $\br \cdot \kt_{loc}$ denote averages over the local random operators 
are of interest, they are harder to compute than one wherein the local operators are independent random matrices $U_{Aj}$, $U_{Bj}$ at different times $j$. In this case analytical results are possible for the case of operator linear
entanglement entropy and corresponding entangling power \cite{Bhargavi2017}, for example
results therein imply an exponential growth of entangling power:
\beq
\begin{split}
\label{eq:rate-ki}
\langle \ep_l(U_{RMT}^{(n)}(\tau)) \rangle_{loc} = &\overline{\ep_l} \left[1-\left(1-\frac{\ep_l(U_{AB}(\tau))}{\overline{\ep_l}} \right)^n\right] \\
 =&\overline{\ep_l} \left[1-\left(1- \frac{C_N}{2} \sin^2(2\tau) \right)^n\right], 
\end{split}
\eeq
with $C_N=N^2(N^2+1)/(N^2-1)^2 \approx 1+ 3/N^2$ and using Eq.~(\ref{eq:RMTavgs}) $\overline{\ep_l} \approx 1-2/N$.
The braces in $U^{(n)}$ indicates that the local operators vary with time and averages are with respect to the circular unitary ensemble (CUE) sampling the space of unitaries in ${\cal H}_{A,B}$ uniformly. 

\begin{figure}
\includegraphics[width=6.5cm]{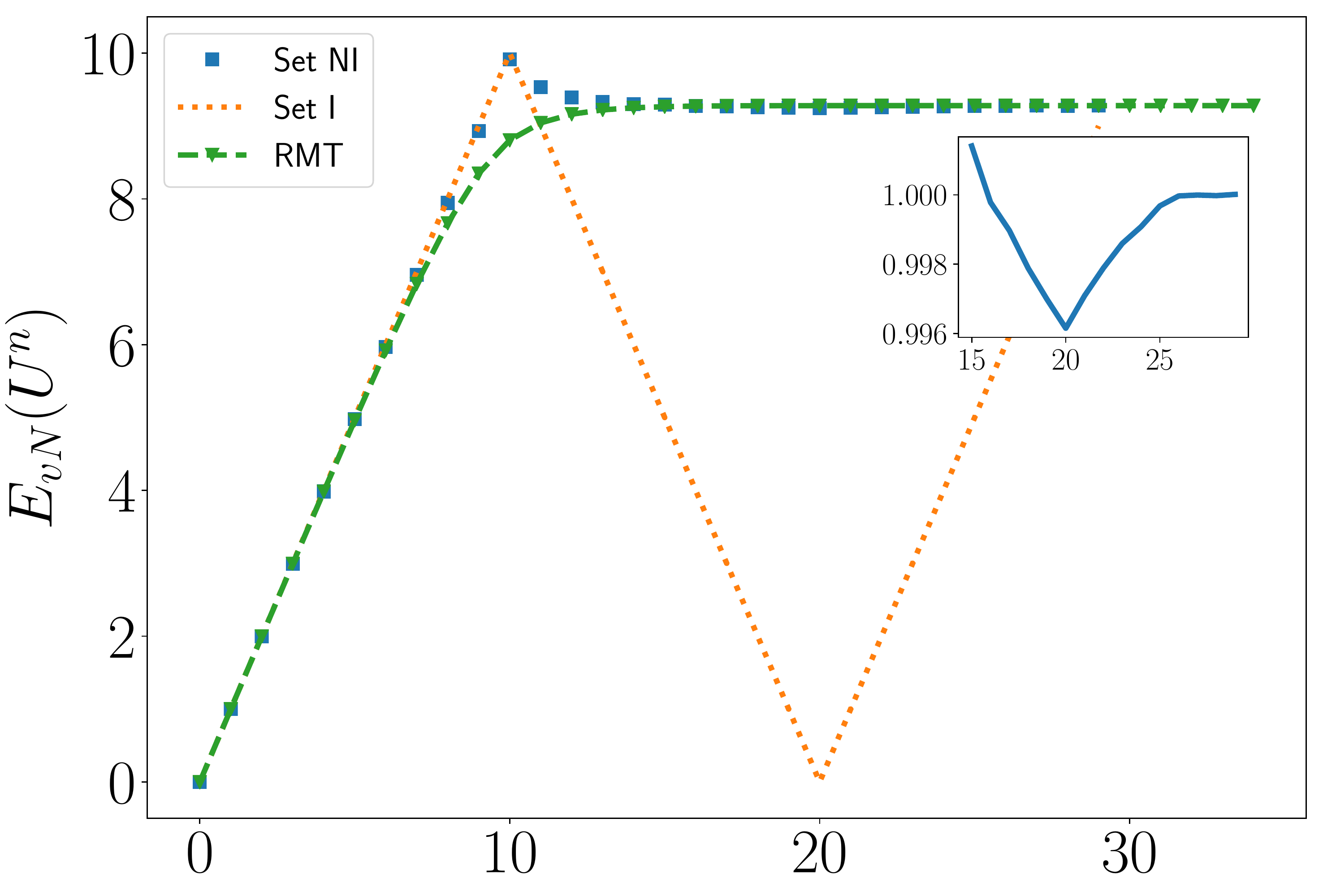}
\includegraphics[width=6.5cm]{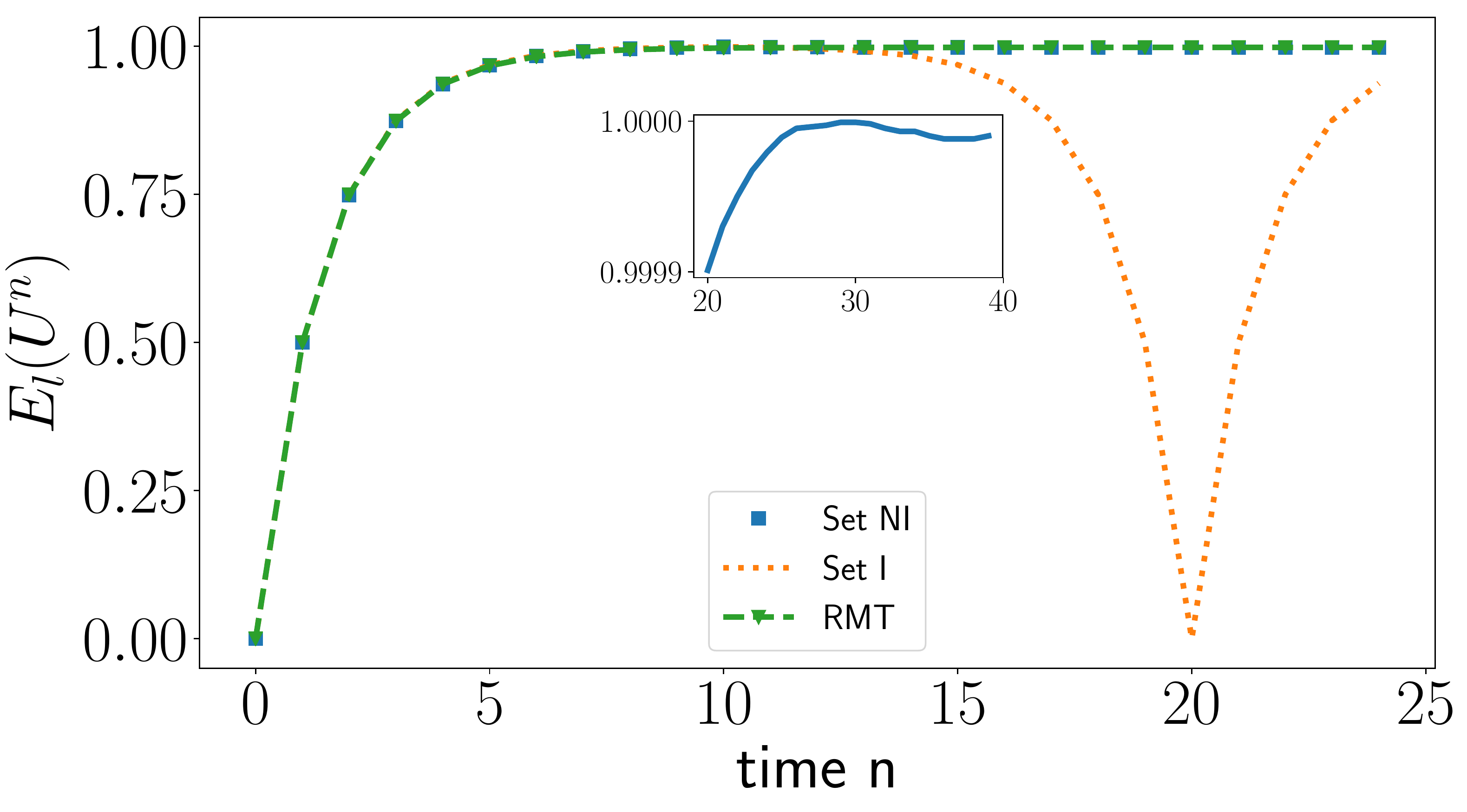}
\caption{Operator von Neumann (top) and linear (bottom) entanglement entropies, $E_{vN,l}(U^n(\tau,h))$ for nonintegrable (Set NI) and integrable cases (Set I), with $\tau=\pi/4$ and $L=10$ spins. Also shown are the corresponding RMT model results.
Inset shows the ratio of the entropies for the non-integrable model  with the corresponding RMT averages from Eq.~(\ref{eq:RMTavgs}). }
\label{fig:EvNlin1}
\end{figure}

We have verified that the averages do not change if we choose, the same CUE's for both $U_{Aj}$ and $U_{Bj}$ for a particular time $j$ or even choose a single CUE for both for all $j$, as is the case with the spin model. 
Thus we can expect these to provide estimates for $U^n(\tau,h)$. There are also similar expressions for averages of $E_l(U^n)$, $E_l(U^n\, S)$ \cite{Bhargavi2017}, which results in 
\beq
\langle E_l(U_{RMT}^{(n)}(\tau)) \rangle_{loc} \approx 1-\left(1- \frac{1}{2} \sin^2(2\tau) \right)^n,
\eeq
where the approximation neglects terms of order $1/N^2$.
Apart from the factor $\overline{\ep_l}$ this is just the entangling power and hence the operator entanglement and entangling power for the RMT model are essentially the same. We note that this already singles out $\tau=\pi/4$ as a case of maximal growth of entangling power, when $\ep_l(U^n)$ and $E_l(U^n)$ can be expected to grow as $1-2^{-n}$ when the magnetic field configurations lead to nonintegrable chains. For measures based on the von Neumann entropy we resort to numerically computing the average over many CUE realizations. It maybe pointed out that the value of $\tau=0.8$ used in \cite{Luitz2017} is in fact very close to $\pi/4$ and we expect and find qualitatively identical results.
%

{\it Operator entanglement entropy vs entangling power:}
The operator entanglement entropies $E_l(U^n)$ and $E_{vN}(U^n)$  as a function of time $n$, the number of kicks, is shown in Fig.~(\ref{fig:EvNlin1}) for $\tau=\pi/4$. These are exactly solvable for the integrable model (Set-I) as the $\lambda_i$ in the Schmidt decomposition of $U^n(\pi/4,1,0,0,)$ are all equal to $1/2^n$ for $0 \le n \le L$. This is most transparent on using Majorana fermions, however see Supplementary material for details of a proof based on Pauli spin operators. During this time we get simply 
\beq
E_l(U^n)=1-2^{-n}, \, E_{vN}(U^n)=n.
\eeq
Remarkably the $E_l(U^n)$ for the integrable chain coincides exactly with that for the RMT model, and thus both increase at the maximum rate. While we do not have a formula for the $E_{vN}$ of the RMT model, numerical simulation in Fig.~(\ref{fig:EvNlin1}) shows that it shares the linear growth for a long time till just before $n=L$ it turns and saturates to the value $\approx 2 \log N-1/(2 \ln2) \approx 9.28$, as $L=10$. Quite suprisingly the nonintegrable model increases just as much as the integrable model does, getting nearly maximally entangled at $n=L$, before disentangling and relaxing to the RMT average.

\begin{figure}
\includegraphics[width=7cm,height=4cm]{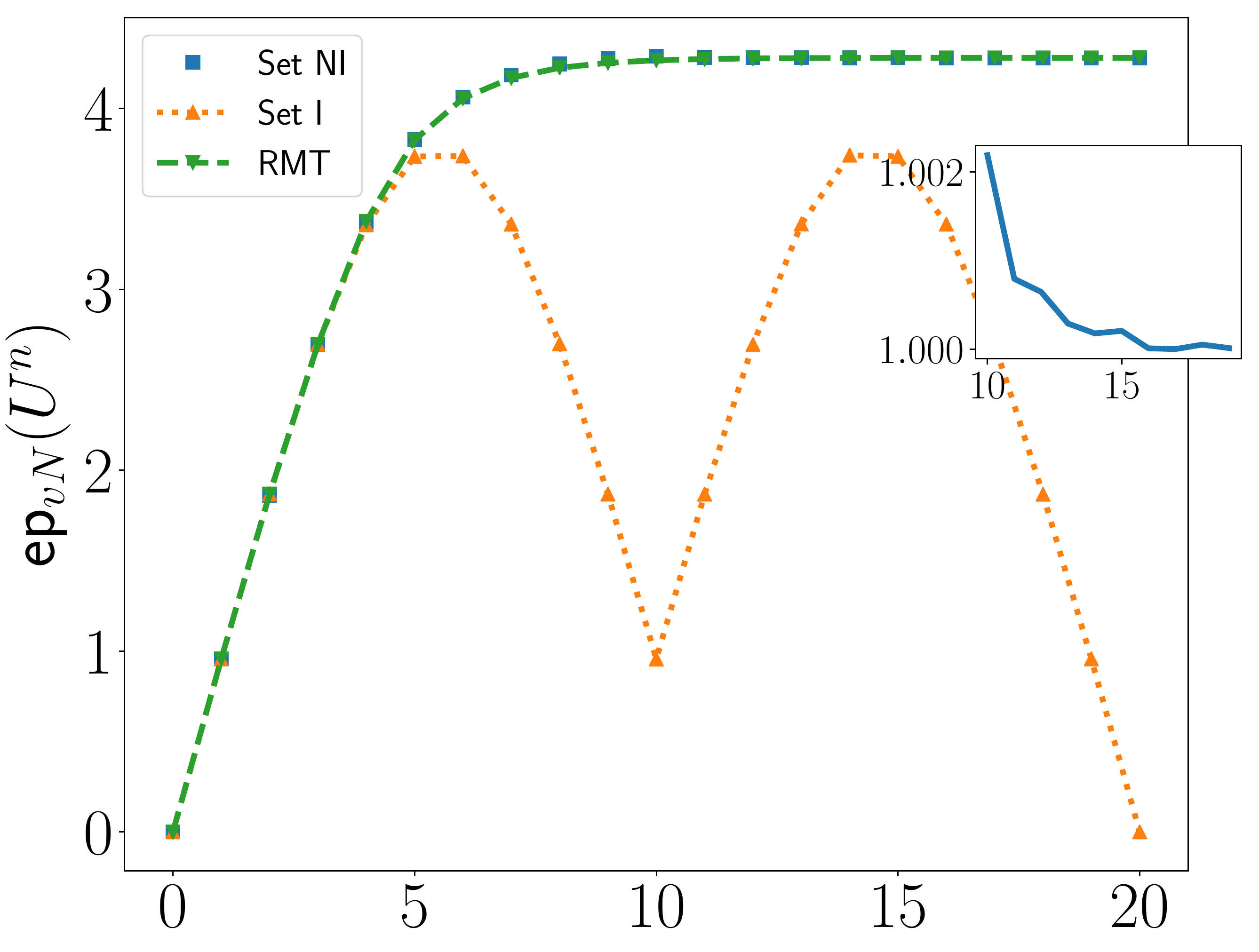}   
  \includegraphics[width=7cm,height=4cm]{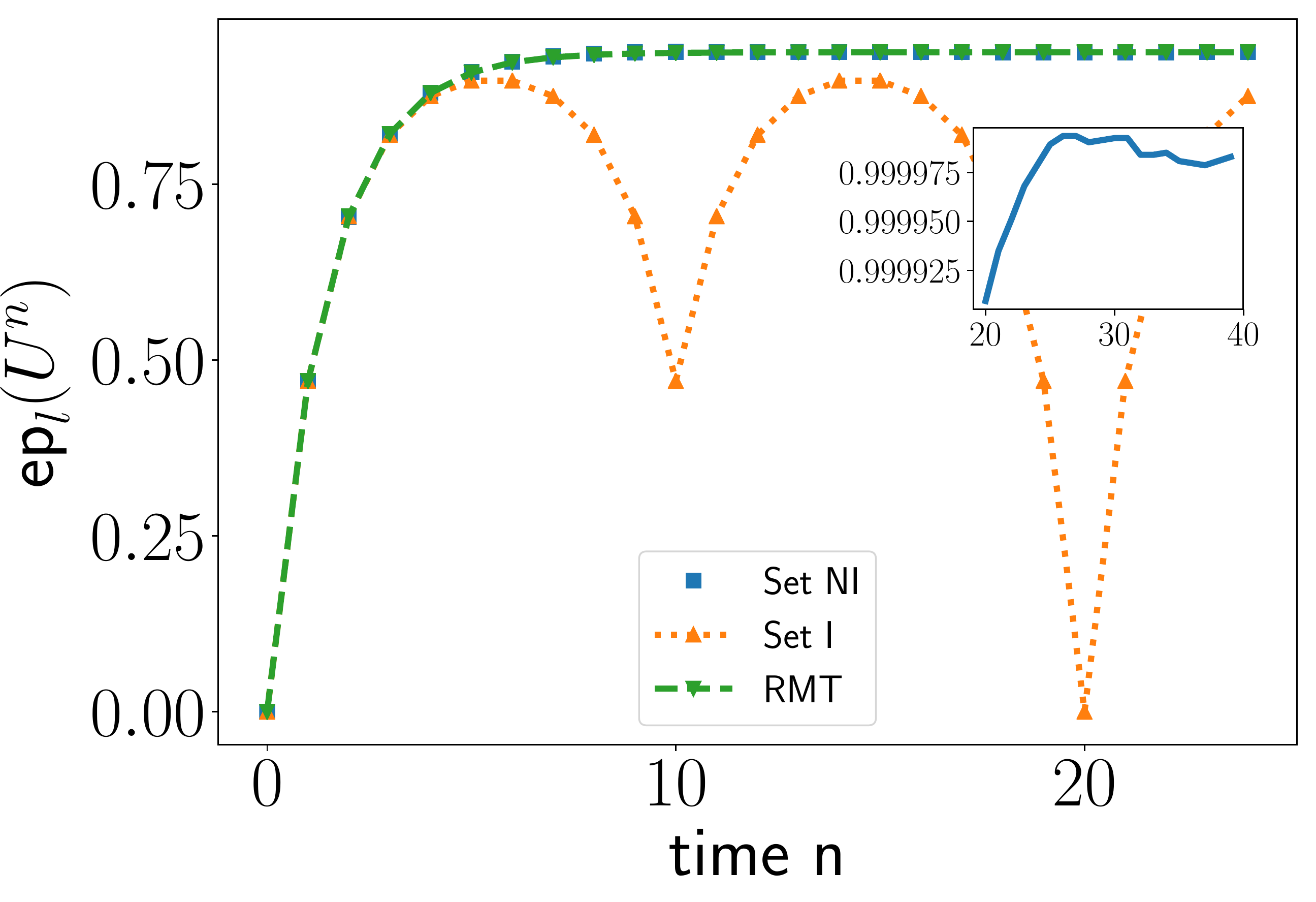}
  \caption{ The  von Neumann (top) and linear (bottom) entangling powers, ${\ep}_{vN,l}(U^n(\tau,h))$ for the nonintegrable (Set-NI) and the integrable cases (Set-I), with $\tau=\pi/4$ and $L=10$ spins. Also shown are the corresponding RMT model results.
  Inset shows the ratio of the entangling powers for the nonintegrable model  with the corresponding RMT averages in Eq.~(\ref{eq:RMTavgs}).}
\label{fig:entpowpiby4}
\end{figure}


\begin{figure}
   \includegraphics[width=7cm,height=4cm]{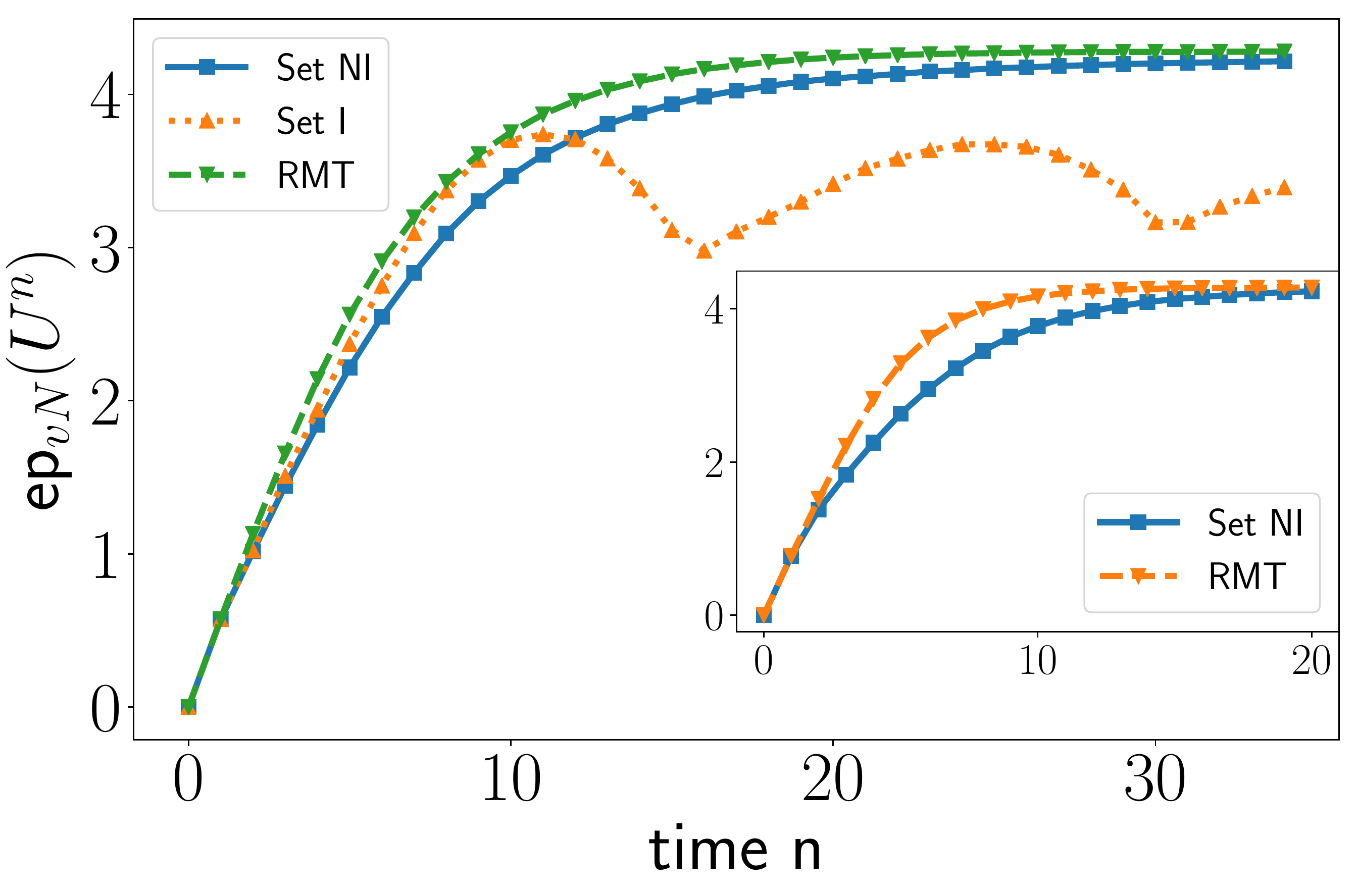}
  \caption{The  von Neumann entropy entangling power, ${\ep}_{vN}(U^n(\tau,h))$ for the nonintegrable Set-NI, integrable Set-I, with $\tau=\pi / 8$ and $L=10$ spins. Also shown are the corresponding RMT model results. Inset shows the same for the nonintegrable and RMT model with $\tau=\pi / 3$. }
\label{fig:entpowpiby8piby3}
\end{figure}

While these results may indicate superior entangling capabilities of the integrable model, except for its complete eventual disentanglement at $n=2L$,
it must be borne in mind that the operator entanglement is the result of acting on a particular pair of maximally entangled states involving ancillas. The entangling power on the other hand is the effect on a  democratic choice of product states. Figure (\ref{fig:entpowpiby4}) shows the variation of $\ep_{l,vN}(U^n(\tau,h))$ for $\tau=\pi/4$.  It is interesting that in contrast to the operator entanglement that continues to increase till $L$ kicks for both the integrable and nonintegrable models, entangling power reaches a maximum at $n=L/2$ for the integrable model and starts decreasing reaching a local minimum after $L$ kicks. On the other hand the nonintegrable model continues to increase during this time and saturates to the RMT value, the contrast being clearer in the von Neumann entropy. The integrable model at $\tau=\pi/4$ can be exactly solved again for the linear entropy entangling power, $\ep_l(U^n)=0$ for $n=0$ and for $1\le n \le L$ it is 
\beq
\ep_l(U^n)=\dfrac{1+2^L -2^{L-n}-2^{n-1}}{(1+2^{L/2})^2}.
\eeq
The Supplementary materials contain a proof of this. Beyond this time, the entangling power is symmetric and becomes zero at $n=2L$.
The maximum entangling power occurs at $n=L/2$ when it is for large $L$ 
$\approx 1-(7/2) 2^{-L/2}$,while the nonintegrable model reaches the RMT average $\overline{\ep_l} \approx 1- (2) \, 2^{-L/2}$.
Also, it is clear from Fig.~(\ref{fig:entpowpiby4}) that the RMT model prediction for the the entangling power in Eq.~(\ref{eq:rate-ki}) works very well for the nonintegrable case at this value of $\tau$. So also the RMT model works for the von Neumman entangling power found numerically and the saturation is very close to that of random states: $\log(N)-1/(2 \ln(2))\approx 4.28$ for $L=10$ spins. Thus in terms of ability to create entanglement on an average, the nonintegrable model in Set-NI can eventually swamp the integrable one in Set-I, even though their operator entanglement entropies grow at the same rate.    

An insight into this difference is provided by the behavior of $E_l(U^n S)$ which decreases considerably for the integrable model compensating for the increasing $E_l(U^n)$. In contrast $E_l(U^n S)$ is nearly a constant for the nonintegrable model. Using the ancilla interpretation of the operator entanglements, this is reflective of the fact that entanglement is shared in a more multipartite manner in the almost random states created by the nonintegrable evolution. See Supplementary material for further details.

For values of $\tau$ different from $\pi/4$ a varied scenario develops 
when comparing the integrable and the nonintegrable.
The entangling power for $\tau=\pi/8$ is shown in Fig.~ (\ref{fig:entpowpiby8piby3}). While the integrable may outstrip the nonintegrable in rate of entangling power, the nonintegrable eventually develops a larger value.
Unlike in the case of $\tau=\pi/4$, in the integrable case this does not vanish and shows fairly small oscillations about what maybe an equilibrium. Also the RMT model predicts a larger power in the growth phase and is not as good as at $\tau=\pi/4$. This maybe attributed to the fact that at $\tau=\pi/8$ the model is not quite as ``chaotic" as at $\tau=\pi/4$. 

\begin{figure}
   \includegraphics[width=7cm,height=4cm]{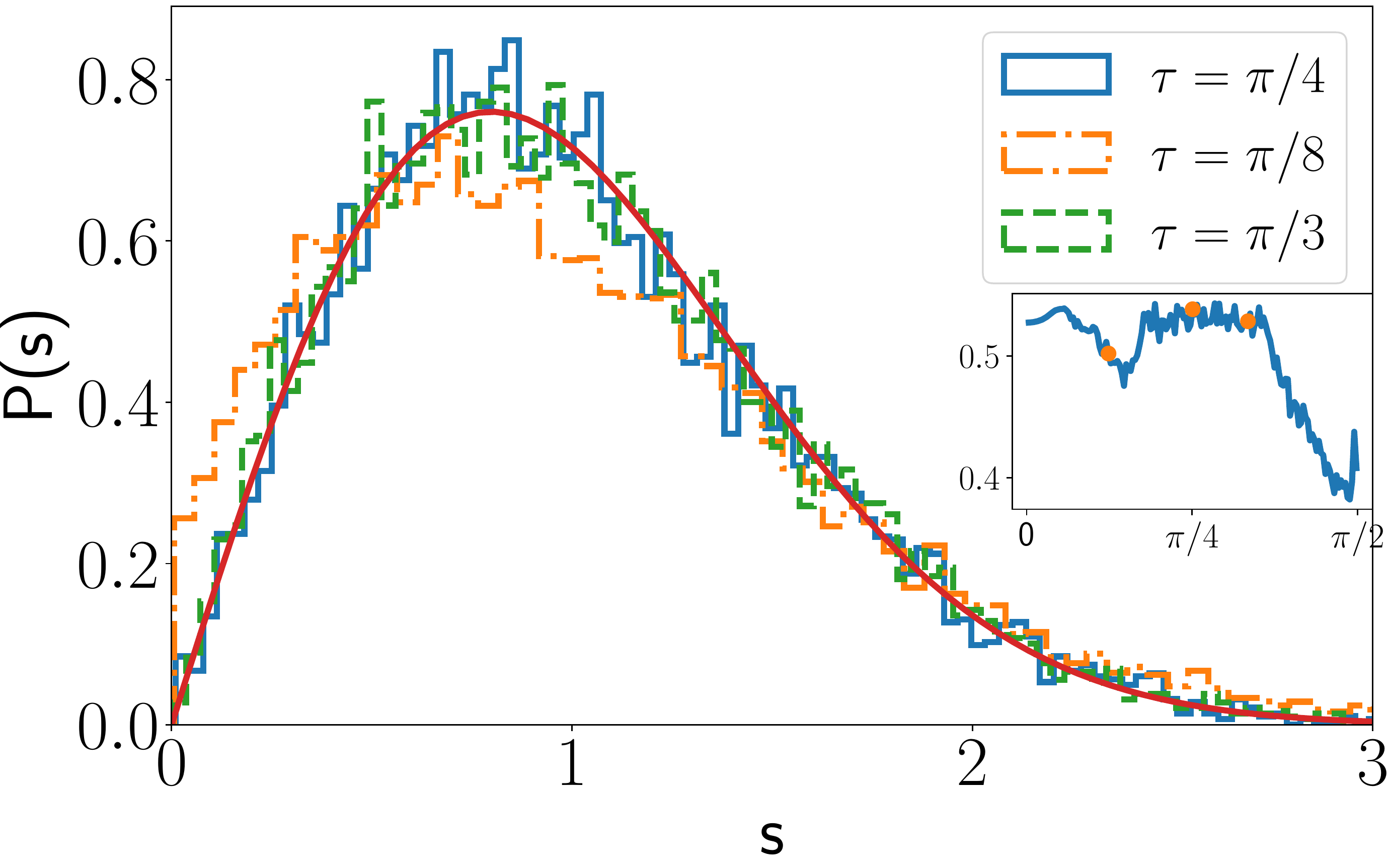}
  \caption{Distribution of spacing for $\tau=\pi/3, \pi/4, \pi /8$ for $L=14$. Inset shows the mean of the ratio of nearest neighbor spacing distribution(NSS) of the operator $U$  {\it{vs}} time between kicks $\tau$.Dots in the inset
  correspond to $\tau=$ $\pi/3$, $\pi/4$ and $\pi/8$.}
\label{fig:rmtprop}
\end{figure}

This is borne out from the NNS distribution and ratio of spacings (using desymmetrized spectrum of even parity states) as illustrated in Fig.~(\ref{fig:rmtprop}), which deviates at $\tau=\pi/4$ from the Wigner distribution.
While the presence of the $\sigma_y$ terms in the Hamiltonian
may suggest time-reversal violation \cite{Luitz2017}, it actually has a false-time-reversal violation and follows the statistics of the orthogonal ensemble COE, see Supplementary material for elucidation.
It also appears that the saturation value of the entangling power and operator entanglement (not shown) is slightly smaller than the RMT value.

However a puzzle arises in the case of $\pi/3$ which seems as RMT-like as the operator at $\pi/4$ in terms of NNS statistics, for example the average ratio of spacings is $\approx 0.53$ in both cases, and its saturation value is that of 
the RMT, yet in its growth phase it deviates significantly from the RMT model values as seen in Fig.~(\ref{fig:entpowpiby8piby3}). 

\begin{figure}
\includegraphics[width=7cm,height=4cm]{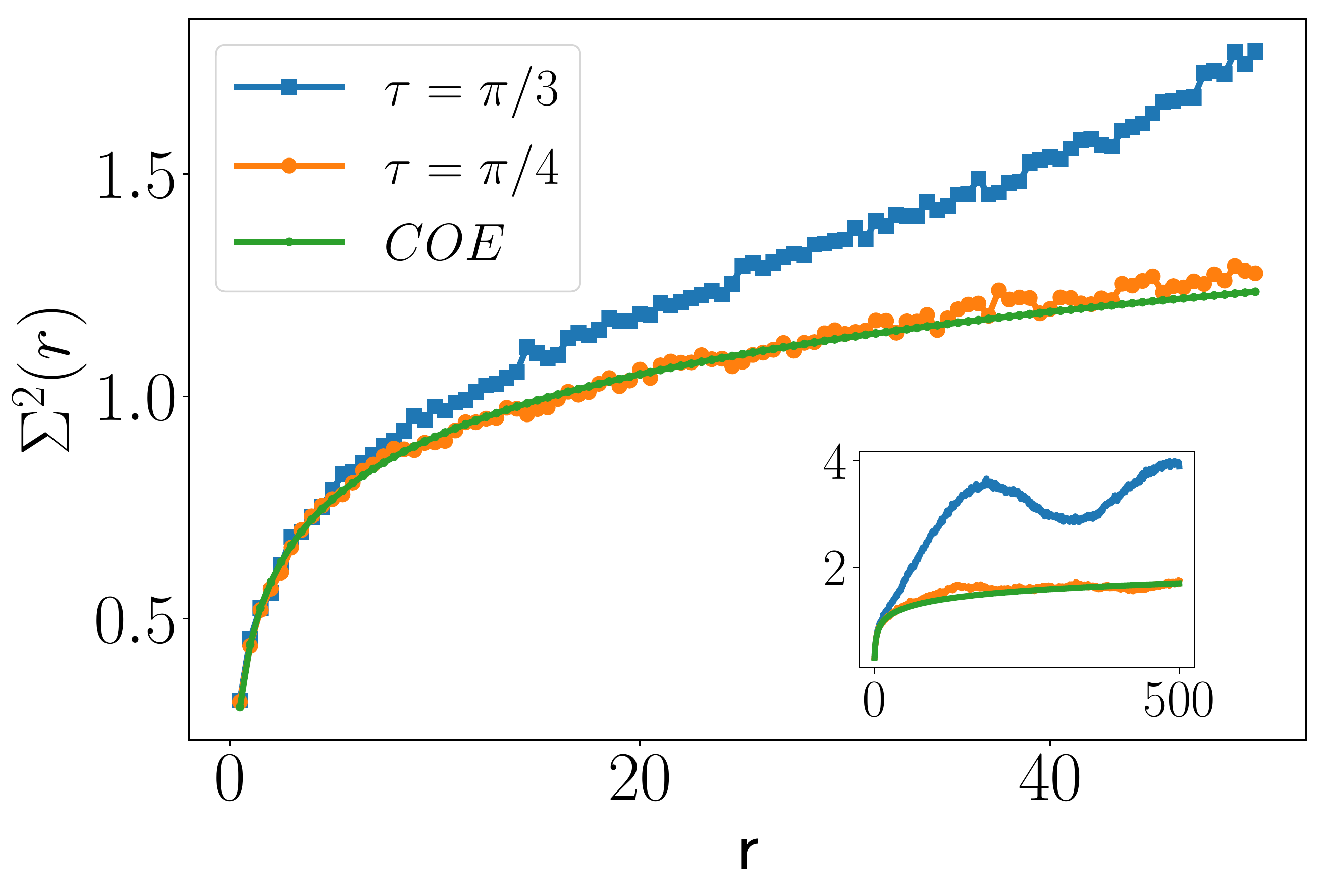}
\caption{Number variance for   $\tau=\pi/3$ and $\tau=\pi/4$. Inset shows the same in a longer range.}
\label{number-var}
\end{figure}

Short time properties are determined by long-range energy correlations \cite{Berry85} than the short-ranged NNS. For example the number-variance \cite{Mehta2004} is one such quantity and is defined as  $\Sigma^2(r)= \langle (n(r) - r)^2 \rangle_{\Delta} $, with $n(r)$ denoting the number of energy levels in an energy window of width $r$ in an unfolded spectrum and $\langle \cdot  \rangle_{\Delta}$ denoting average over length $r$ windows. Figure~\ref{number-var} shows this quantity as a function of $r$ for  $\tau=\pi/3$ and $\tau=\pi/4$. While, they match for about $10$ mean spacings, they deviate thereafter, with $U(\pi/4)$ following the RMT number variance for a much longer scale, while the other case is intermediate to the Poisson or integrable limit. Thus the study of long-range statistics in many-body systems may be crucial to distinguish those that are not full chaotic. Unlike systems with a classical limit these many-body systems are yet to be classified in terms of the extent of the chaos present.  

\begin{figure}
\includegraphics[width=6.5cm,height=3.5cm]{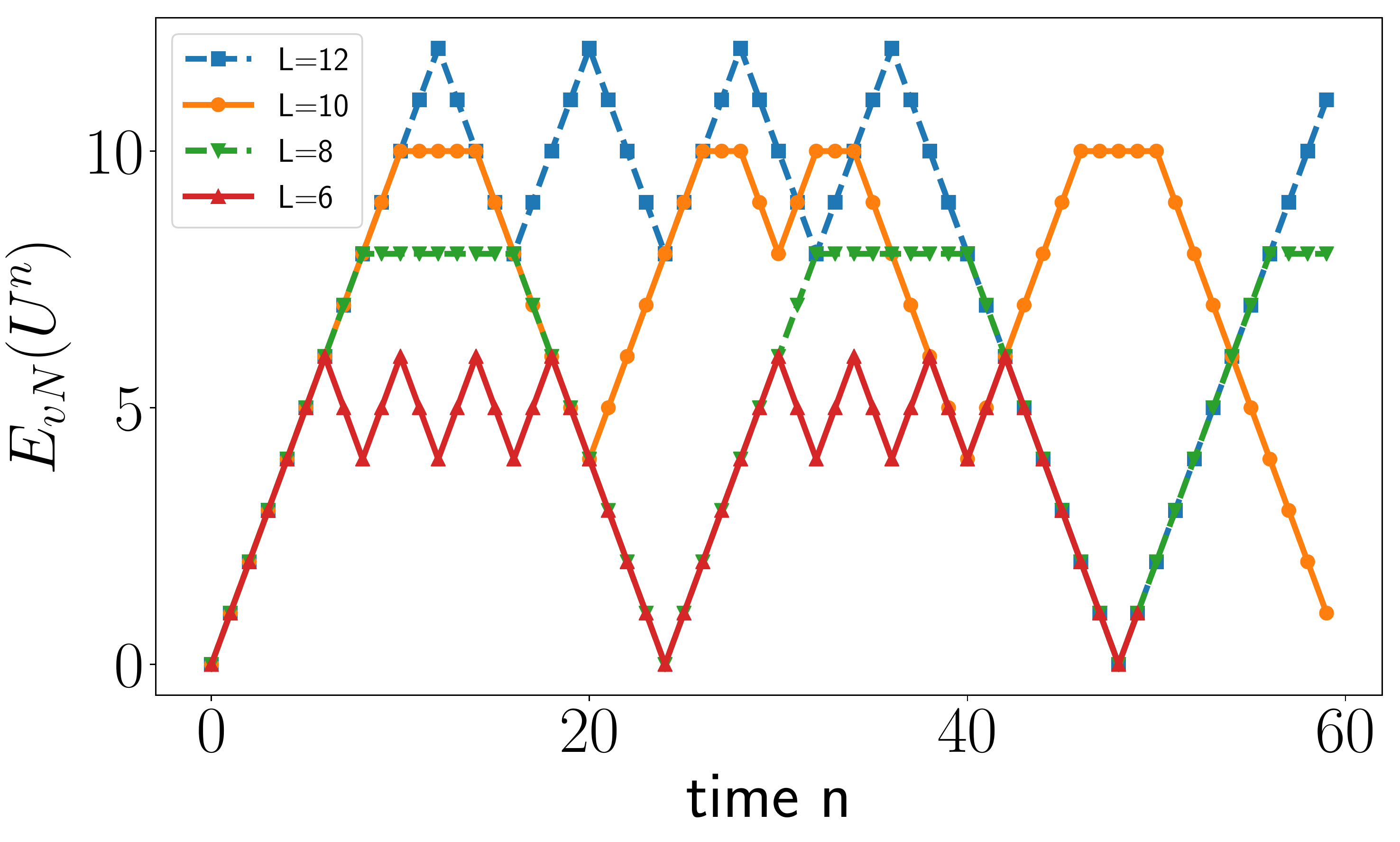}
\includegraphics[width=6.5cm,height=3.5cm]{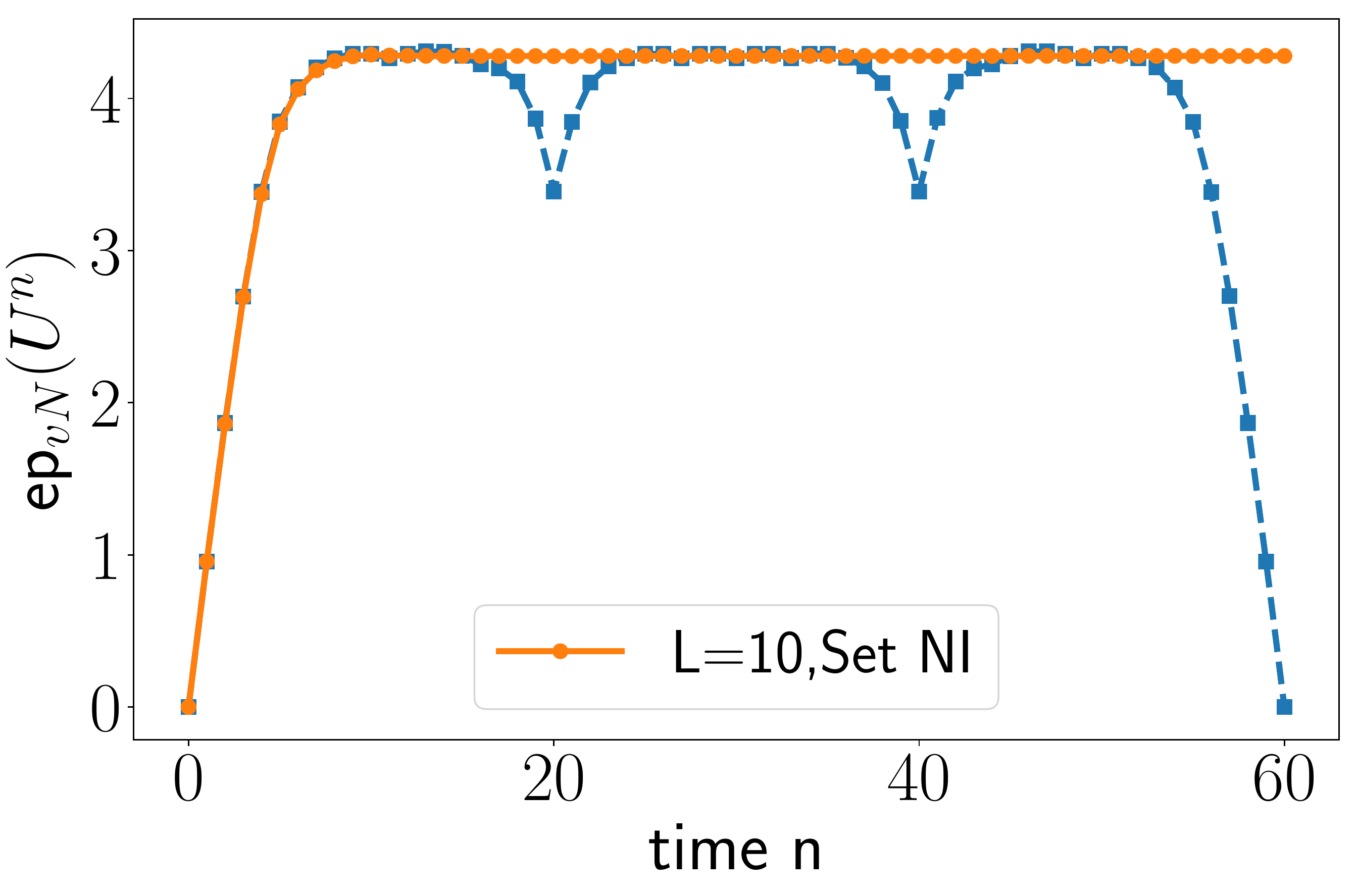}
\caption{Operator von Neumann entropy (top) $E_{vN}(U^n(\tau,h))$ and entangling power (bottom) when $h_x^i=1.0,h_y^i=0,h_z^i=1$, and $\tau=\pi/4$. The nonintegrable case Set-NI is shown in the latter figure for comparison.}
\label{set3evn}
\end{figure}

Finally we display a case that has features of both the integrable and nonintegrable models, when for all $i$: $h_x^i=1.0,h_y^i=0,h_z^i=1$ and $\tau=\pi/4$. Figure~\ref{set3evn} shows the variation of operator entanglement entropy with the number of kicks in this model. However, the entangling power even if periodic, reaches as high a value as that of the nonintegrable case. Being an interacting model it promises to be an interesting one for further studies.

Relying on analytical and numerical RMT averages for bipartite entangling powers we have seen that while it sometimes describes very well the time evolution of such measures, 
it mostly provides an upper bound for other nonintegrable situations. Thus one outstanding work is to find entangling power in random circuits that take into account the internal structure of
local blocks and may provide better estimates in other cases. While the measures studied in this paper are entanglement based ones, connections to other measures that are being intensively pursued,
such as operator spreading \cite{Sondhi18} and information scrambling are to be explored, as also multipartite measures of entangling power.

\section{Acknowledgements}
R.P. would like to gratefully acknowledge Ravi Prakash for useful discussions and help with computing the number variance.

\bibliographystyle{unsrt}
\bibliography{absnew}
\onecolumngrid

\newpage

\section{Supplementary Material}

\section{Comparison of  $E_l(U^n)$ and $E_l(U^nS)$ for Set-I and Set-NI}

\begin{figure}[!h]
\includegraphics[width=0.7\linewidth]{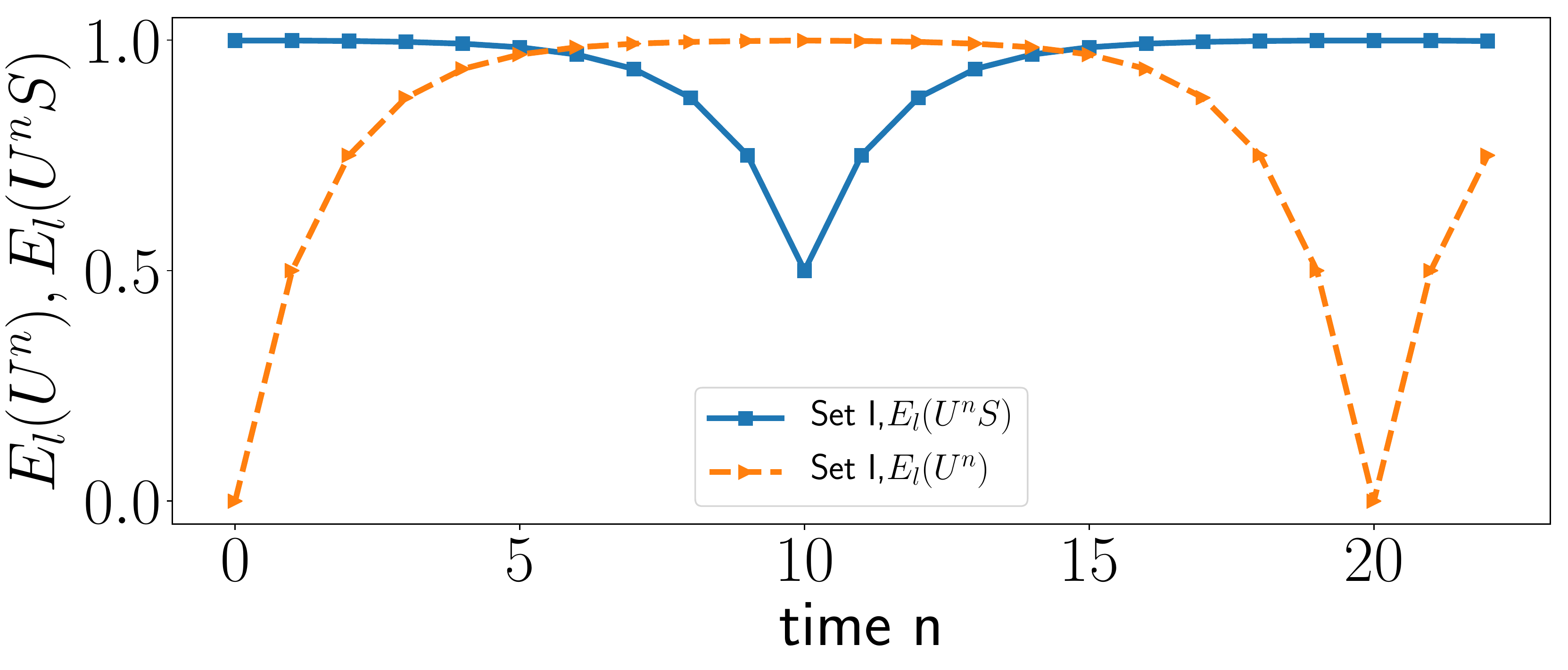}
\caption{Linear entropy operator entanglements  $E_{l}(U^n(\tau,h))$ and $E_{l}(U^n(\tau,h)S)$ for the integrable case (Set-I), with $\tau=\pi/4$ and $L=10$ spins.}
\label{fig:Ilin1}
\end{figure}

\begin{figure}[!h]
\includegraphics[width=0.7\linewidth]{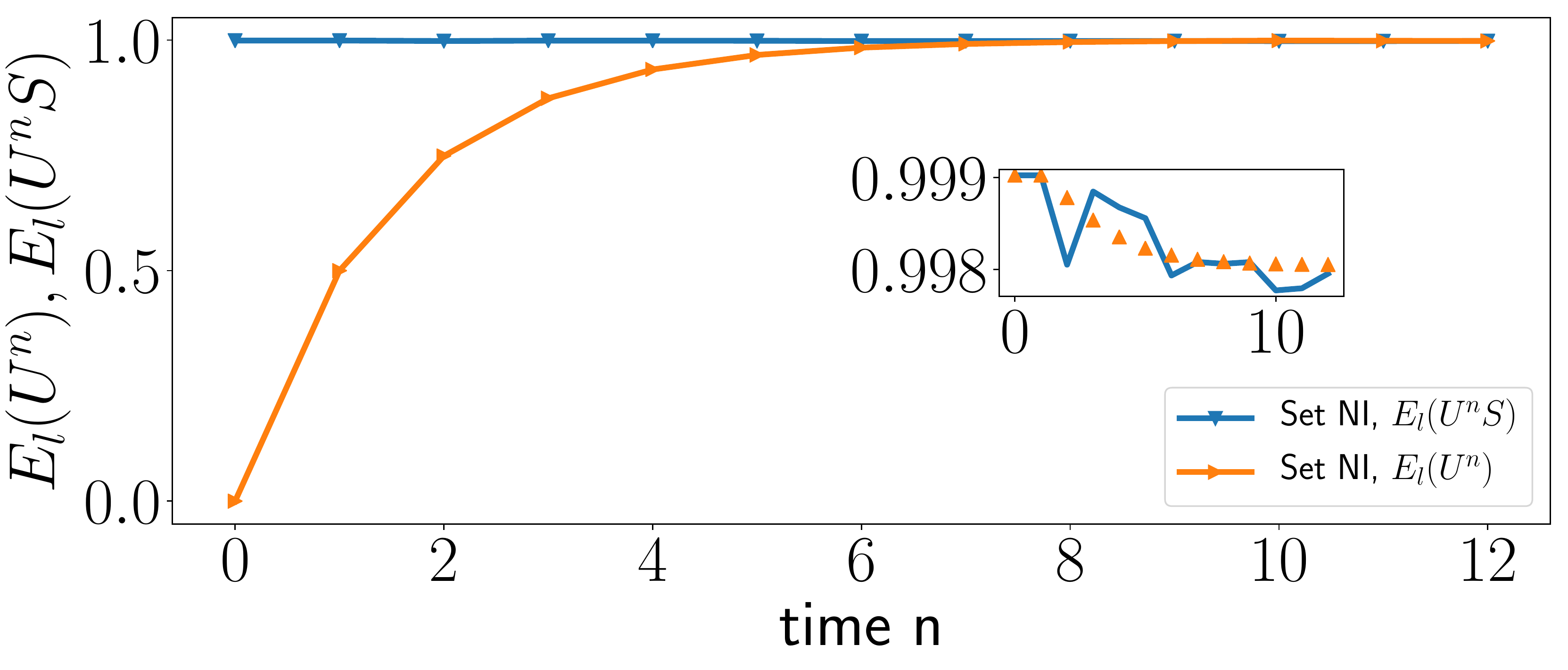}
\caption{Linear entropy operator entanglements $E_{l}(U^n(\tau,h))$ and $E_{l}(U^n(\tau,h)S)$ for the non-integrable case (Set-NI), with $\tau=\pi/4$ and $L=10$ spins. Inset shows details of $E_{l}(U^n(\tau,h)S)$ for the  non-integrable case and also the RMT model prediction (triangles).}
\label{fig:NIlin1}
\end{figure}

Figures (\ref{fig:Ilin1}) and (\ref{fig:NIlin1}) show variation of  $E_l(U^n)$ and $E_l(U^nS)$ with time  respectively for the integrable (Set I) and the non-integrable (Set NI) case studied in this paper.  As mentioned in the main letter we see that
 $E_l(U^nS)$ remains almost close to the maximum value it starts with for the non-integrable case. In the integrable case on the other hand, it dips to a minimum value of $1/2$, at time $n=L$. This can be understood from the fact that the non-integrable 
 evolution leads to a random $4$-party state in the ancilla picure (see main letter) which has near maximal entanglement in different bipartitions.  The integrable model on the other hand, in the ancilla picure leads to a state which has  maximal entanglement across 
the $AA'|BB'$ cut, but little entanglement across the  $AB'|AB'$ cut. These features are qualitatively seen to hold for the von Neumann entropy as well, even though we do not have a similar relation between von Neumann entangling power and 
operator von Neumann entropy.  

Throughout the rest of the supplementary material, except in the part of the antiunitary symmetry. we consider the integrable kicked Ising model, with the parameters in Set-I $(h^x_i=1.\, 0,h^y_i=0.0,\, h^z_i=0)$ and time between kicks $\tau=\pi/4$. The time evolution or Floquet operator, we recall, is thus given by,
\begin{equation}
U= \exp\left(-i \frac{\pi}{4} \sum_{j=1}^L \sigma_j^z \sigma_{j+1}^z \right) \exp\left(-i \frac{\pi}{4} \sum_{j=1}^L \sigma_j^x \right).  
\end{equation}

\section{Nonlocal part of $U^n$}

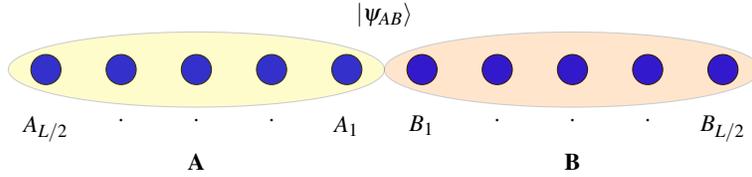
\begin{figure}[h]
\begin{tikzpicture}
\draw [fill=blue] (1,1) circle [radius=.2];\draw [fill=blue] (2,1) circle [radius=.2];\draw [fill=blue] (3,1) circle [radius=.2];\draw [fill=blue] (4,1) circle [radius=.2];\draw [fill=blue] (5,1) circle [radius=.2];\draw [fill=blue] (6,1) circle [radius=.2];\draw [fill=blue] (7,1) circle [radius=.2];\draw [fill=blue] (8,1) circle [radius=.2];\draw [fill=blue] (9,1) circle [radius=.2];\draw [fill=blue] (10,1) circle [radius=.2]; \draw[fill =yellow, opacity=.2] (3,1) ellipse [x radius =2.5, y radius =.5]; \draw[fill =orange, opacity=.2] (8,1) ellipse [x radius =2.5, y radius =.5] node [below,opacity=1] at (3,0) {{\bf A}} node [below,opacity=1] at (8,0) {{\bf B}} node [below,opacity=1] at (5.5,2.0) {$|\psi_{AB}\kt $} node [below,opacity=1] at (5,.5) {$A_1$}  node [below,opacity=1] at (4,.5) {$ \cdot $}  node [below,opacity=1] at (3,.5) {$\cdot$}  node [below,opacity=1] at (2,.5) {$\cdot$}  node [below,opacity=1] at (1,.5) {$A_{L/2}$} node [below,opacity=1] at (6,.5) {$B_1$}  node [below,opacity=1] at (7,.5) {$ \cdot $}  node [below,opacity=1] at (8,.5) {$\cdot$}  node [below,opacity=1] at (9,.5) {$\cdot$}  node [below,opacity=1] at (10,.5) {$B_{L/2}$};
\end{tikzpicture}
\caption{Schematic of the labeling scheme used for the two partitions $A$ and $B$.}
\label{fig:spinfigure}
\end{figure}

Introduce  the following notation for spin operators of different partitions, for $j \leq M=L/2$,  
\begin{equation}
\vec{A_j} \equiv \vec{\sigma}_{M+1-j},  \text{ and }   \vec{B_j} \equiv  \vec{\sigma}_{M+j}.
\end{equation}
Thus $A_1$ and $B_1$ represent Pauli matrices for spins at locations $M$ and $M+1$ respectively, see Fig.~(\ref{fig:spinfigure}).

Local operators are those that belong exclusively to the space of $A$ or $B$ spins. It was shown in \cite{LSM15} that the {\it nonlocal} part of $U^n$ is $(U^n)_{nl}=\prod_{i=1}^nV_i $, where the mutually commuting operators $V_i$ are given by,
\begin{equation}
\label{nleLby2}
V_i=\frac{1}{\sqrt{2}} \left (I-iA_i^yB_i^y \prod_{j=1}^{i-1}A_j^xB_j^x \right), \;\;\; 1 \le i \le L/2.  
\end{equation}
Beyond $i=L / 2$ we have,
\begin{equation}
\label{ngeLby2}
V_{\frac{L}{2}+k}=\frac{1}{\sqrt{2}} \left(I-iA_{\frac{L}{2}-k+1}^zB_{\frac{L}{2}-k+1}^z \prod_{j=1}^{\frac{L}{2}-k}A_j^xB_j^x \right), \;\;\;  1 \le k \le L/2,\; \text{ and }, \; V_{L+k}=V_k.
\end{equation}

It is also easy to see that $V_j^2$ is a local operator as far as the $AB$ partition is concerned as 
\begin{equation}
V_j^2 =-i A_j^y \prod_{k=1}^{j-1} A_k^x\, \ot \, B_j^y \prod_{k=1}^{j-1} B_k^x
\end{equation}
Note that $V_1$ and $V_L$ involve only the $A_1^y, B_1^y$ and $A_1^z, B_1^z$ operators respectively. The $V_i$ contain precisely strings of operators that appear in the Jordan-Wigner transform of spins to Majorana fermions. Thus although we can interpret the results elegantly in terms of entanglement between Majorana fermions of two species (those in $A$ and $B$), we proceed with the spin language as it provides the Schmidt decompositions of operators written with spin variables. 
All the measures used in this work, $E_{l,vN}(U)$, $E_{l,vN}(US)$, $\ep_{l,vN}(U)$ are local operator invariants(\cite{Bhargavi2017}), that is they are the same as for $(U_A\otimes U_B) \, U \, (U_A'\otimes U_B')$. They can hence be obtained by just considering the nonlocal part of $U^n \equiv (U^n)_{nl}$.

\section{Proofs for operator entanglements: $E_l(U^n)=1-2^{-n}$, $E_{vN}(U^n)=n$ }

In this section, we find how the linear and von Neumann operator entanglement entropies grow with the number of kicks $n$. We do this by showing that
as we multiply the operators $V_i$ (Eqn.~(\ref{nleLby2})), in each step we get an operator Schmidt decomposition  with equal
Schmidt coefficients, with the Schmidt rank simply doubling at each step, till $n \leq L/2$. Thus, $(U^n)_{nl}$ is 
an operator with Schmidt rank $2^n$, with equal Schmidt coefficients. 

For, $i > L/2$ the structure of $V_i$ changes, as in Eq.~(\ref{ngeLby2}). Suppose, we are interested in $n=L/2+m$, the strategy is to define $m$ operators, $V'_{m+1-k}=V_{\frac{L}{2}+k}V_{\frac{L}{2}-k+1}, 1 \le k \leq m$. We already know that, $\prod_{i=1}^{L/2-m} V_i$ is a maximally entangled operator with Schmidt rank $2^{(L/2-m)}$. We will then show, taking advantage of the commutativity of $V_j$, that multiplying with a $V'$ quadrupules the Schmidt rank so that  $(U^{\frac{L}{2}+m})_{nl}$ is a
operator with  Schmidt rank  $2^{L/2-m} 2^{2m}=2^n$. This immediately leads to the desired result.

\vspace{10mm}
{\textbf{Theorem 1:}} $E_l(U^n)=1-2^{-n}, \; \; E_{vN}(U^n)=n, \; \; E_{l,vN}(U^{2L-n})=E_{l,vN}(U^n)$, $1 \leq n \leq L$. 

\vspace{5mm}
{\textbf{Proof:}}

First assume that  $n \leq L/2$. To begin with notice by direct computation that 
\begin{equation}
V_1= \frac{1}{\sqrt{2}} (I_1\otimes I_1-iA_1^y \otimes B_1^y), \;
V_2V_1= \frac{1}{2}(I_{12}\otimes I_{12}-iA^y_1 \ot B^y_1-iA^x_1A^y_2 \ot B^x_1B^y_2 + A^z_1A^y_2 \ot B^z_1B^y_2), 
\end{equation}
are already in the Schmidt decomposed form except for the phases that can be absorbed into the operators. They are maximally entangled with 
entropies $1$ and $2$ respectively as the Schmidt coefficients are $(1/\sqrt{2},1/\sqrt{2})$, and $(1/2,1/2,1/2,1/2)$.
To proceed by induction assume that 
\begin{equation}
\prod_{i=1}^{n-1}V_i = \frac{1}{2^{(n-1)/2}} \sum_{j=1}^{{2^{(n-1)}}} \ca_j^{(n-1)} \otimes \cb_j^{(n-1)}, \; \mbox{with} \;\Tr(
\ca_i^{(n-1) \dagger}\ca_j^{(n-1)})=\Tr(\cb_i^{(n-1)\dagger}\cb_j^{(n-1)})=2^{(n-1)}\delta_{ij}. 
\end{equation} 
Now,
\begin{equation}
V_n= \frac{1}{\sqrt{2}}(I-iA^y_n A_{n-1} \ot B^y_n B_{n-1}),
\end{equation}
with $A_{n-1}=\prod_{k=1}^{n-1}A^x_k $, $B_{n-1}=\prod_{k=1}^{n-1}B^x_k $. 
Thus, 
\[ V_n \, \prod_{i=1}^{n-1}V_i= \frac{1}{2^{n/2}} \sum_{j=1}^{2^{n-1}}\left( \ca_j^{(n-1)} \ot \cb_j^{(n-1)}-i(A^y_n A_{n-1} \ca_j^{(n-1)}) \ot (B^y_n B_{n-1}\cb_j^{(n-1)}) \right).\] 

This is again Schmdit decomposed but for phases, as  $A^y_n A_{n-1} \ca_j^{(n-1)}$ is orthogonal to  $A^y_n A_{n-1} \ca_k^{(n-1)}$, 
for all $j \ne k$ (as $A_n^y A_{n-1}$ is unitary)
and also $A^y_n A_{n-1} \ca_j^{(n-1)}$ is orthogonal to $\ca _k^{(n-1)}$, for $j \ne k$ as there is one extra spin from site $n$ in the former. Identical considerations hold for $B$ operators. Hence with suitable redefinitions we again have an orthogonal decomposition,
\begin{equation}
\prod_{i=1}^{n}V_i= \frac{1}{2^{n/2}}\sum_{j=1}^{{2^n}} \ca_j^{(n)} \ot \cb_j^{(n)}. 
\end{equation}
With the normalization of the $\ca^{(n)}$ operators as $\Tr(\ca^{(n) \dagger}\ca^{(n)})=2^{n}$, it follows that the $2^n$ Schmidt coefficients $\lambda_j$ are all $2^{-n}$. Thus the result follows.




For $n \geq L/2$ By virtue of Eq.~(\ref{nleLby2}),
\begin{equation}
V_{\frac{L}{2}-k+1} =  \frac{1}{\sqrt{2}}\left(I-iA_{\frac{L}{2}-k+1}^yB_{\frac{L}{2}-k+1}^y \prod_{j=1}^{\frac{L}{2}-k}A_j^xB_j^x \right), \;\;1 \leq k \le L/2.
\end{equation}

Suppose we are interested in $E(U^{\frac{L}{2}+m}),(1 \leq m \leq L/2)$.  Define $m$ paired up operators as follows
\begin{equation}
\label{V'struc}
V'_{m+1-k}=V_{\frac{L}{2}-k+1}V_{\frac{L}{2}+k} = \frac{1}{2}\left(I-iA^z_{\frac{L}{2}-k+1} \, A_{L/2-k}\,  B^z_{\frac{L}{2}-k+1} \, B_{L/2-k} -i A^y_{\frac{L}{2}-k+1}\,  A_{L/2-k} B^y_{\frac{L}{2}-k+1}\, B_{L/2-k} +  A^x_{\frac{L}{2}-k+1}   B^x_{\frac{L}{2}-k+1} \right), 
\end{equation}
with $1 \le k \le m$ and $A_{L/2-k}=\prod_{j=1}^{\frac{L}{2}-k}A^x_j$ and $B_{L/2-k}$ is similarly defined.
Hence we have, 
\begin{equation}
E_{l,vN}\left(U^{\frac{L}{2}+m}\right) = E_{l,vN} \left(\prod_{j=1}^{\frac{L}{2}-m} V_j \prod_{l=1}^{m} V'_l \right).  
\end{equation}
From the first part of the proof it is clear that $\prod_{j=1}^{\frac{L}{2}-m} V_j$ will have a Schmidt decomposition $\frac{1}{2^{\frac{n}{2}}}\sum_{j=1}^{{2^n}} \ca^{(n)}_j \ot \cb^{(n)}_j$ of rank $2^{n}$ with $n=\frac{L}{2}-m$.  Multiplying these with $V_j'$ operators increases the Schmidt rank $4-$fold each time, each contain $4$ orthogonal terms with a new spin operator at each step. Thus this follows on similar lines as for $U^n$ with $n <L/2$, and we get that the Schmidt rank of $U^{L/2+m}$ is $2^{L/2-m}\times 4^m=2^{L/2+m}$. Hence we have till $n=L$, $E_l(U^n)=1-2^{-n}, \, E_{vN}(U^n)=n$. 

Beyond $n=L$, the operator gets ``disentangled" in a symmetric manner, that is $E_{l,vN}(U^{2L-n})=E_{l,vN}(U^n)$. This follows as $U^{2L}$ is local across the $A|B$ partition, as it's nonlocal part is  
\beq
V_1^2 \cdots V_L^2,
\eeq
and each $V_j^2$ are local. Thus $E(U^{2L-m}=E(U^{-m})=E(U^m)$, the last equality following from the fact that the Schmidt decomposition of an operator and its adjoint only differ in the Schmidt operators being self-adjoints of each other.
Note that $E_{l,vN}(U^{2L})=0$ and the operator gets fully disentangled. 

\section{Proof of $\ep_l(U^n)=\dfrac{1+2^L -2^{L-n}-2^{n-1}}{(1+2^{L/2})^2}$}  

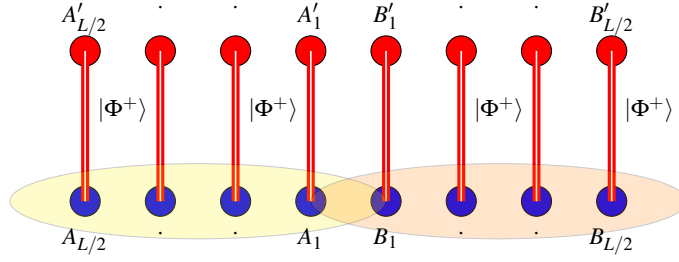
\begin{figure}  
\begin{tikzpicture}
\draw [fill=blue] (1,0) circle [radius=.2];\draw [fill=red] (1,2) circle [radius=.2];\draw[red,very thick,double] (1,0) -- (1,2);\draw [fill=blue] (-1,0) circle [radius=.2];\draw [fill=red] (-1,2) circle [radius=.2];\draw[red,very thick,double] (-1,0) -- (-1,2);\draw [fill=blue] (3,0) circle [radius=.2];\draw [fill=red] (3,2) circle [radius=.2];\draw[red,very thick,double] (3,0) -- (3,2);\draw [fill=blue] (5,0) circle [radius=.2];\draw [fill=red] (5,2) circle [radius=.2];\draw[red,very thick,double] (5,0) -- (5,2);\draw [fill=blue] (0,0) circle [radius=.2];\draw [fill=red] (0,2) circle [radius=.2];\draw[red,very thick,double] (0,0) -- (0,2);\draw [fill=blue] (2,0) circle [radius=.2];\draw [fill=red] (2,2) circle [radius=.2];\draw[red,very thick,double] (2,0) -- (2,2);\draw [fill=blue] (4,0) circle [radius=.2];\draw [fill=red] (4,2) circle [radius=.2];\draw[red,very thick,double] (4,0) -- (4,2);\draw [fill=blue] (6,0) circle [radius=.2];\draw [fill=red] (6,2) circle [radius=.2];\draw[red,very thick,double] (6,0) -- (6,2);\draw[fill =yellow, opacity=.2] (.5,0) ellipse [x radius =2.5, y radius =.5]; \draw[fill =orange, opacity=.2] (4.5,0) ellipse [x radius =2.5, y radius =.5]  
node [below,opacity=1,black] at (2,-.25) {$A_1$}  node [below,opacity=1,black] at (1,-.25) {$ \cdot $}  node [below,opacity=1,black] at (0,-.25) {$\cdot$}  node [below,opacity=1,black] at (-1,-.25) {$A_{L/2}$}  node [below,opacity=1,black] at (3,-.25) {$B_{1}$} node [below,opacity=1,black] at (4,-.25) {$\cdot$}  node [below,opacity=1,black] at (5,-.25) {$ \cdot$}  node [below,opacity=1,black] at (6,-.25) {$B_{L/2}$}  
node [below,opacity=1,black] at (2,2.75) {$A'_1$}  node [below,opacity=1,black] at (1,2.75) {$ \cdot $}  node [below,opacity=1,black] at (0,2.75) {$\cdot$}  node [below,opacity=1,black] at (-1,2.75) {$A'_{L/2}$}  node [below,opacity=1,black] at (3,2.75) {$B'_{1}$} node [below,opacity=1,black] at (4,2.75) {$\cdot$}  node [below,opacity=1,black] at (5,2.75) {$ \cdot$}  node [below,opacity=1,black] at (6,2.75) {$B'_{L/2}$}  
node [below,opacity=1,black] at (6.5,1.5) {$|\Phi^+ \kt$} node [below,opacity=1,black] at (-.5,1.5) {$|\Phi^+ \kt$}node [below,opacity=1,black] at (1.5,1.5) {$|\Phi^+ \kt$}node [below,opacity=1,black] at (4.5,1.5) {$|\Phi^+ \kt$}
;
\end{tikzpicture}
\caption{Ancilla picture for the spin chain: the spins in the lower row are subjected to the spin chain dynamics while the upper (ancilla) are not. Initially corresponding pairs of spins between the  chain and the ancilla are maximally entangled in a Bell state $|\Phi^+\kt.$ 
To find the operator entanglement of $U^nS$, one needs to find the entanglement  across the $A_1A'_1..A_{L/2}A'_{L/2}$ and  $B_1B'_1..B_{L/2}B'_{L/2}$ bipartition  
of the state obtained by the action of $U^nS$ on the $A_1A_2..A_{L/2}B_1B_2..B_{L/2}$ subsystem.}
\label{fig:ancilla}
\end{figure}

Our starting point is the relation, (\cite{zanardi2001})
\begin{equation}
\label{epu}
\ep_l(U)= \frac{N^2}{(N+1)^2} (E_l(U) + E_l(US) - E_l(S)). 
\end{equation}
As we already know, $ E_l(U^n)$ from the previous section we need  $E_l(U^nS)$.  In order to find $E_l(U^nS)$ we will work in the ancilla picture mentioned in the beginning of the main letter as the combinatorics involved is easier to see in terms of states rather than operators. The ancilla picture would require us to consider, a four party state with the dimension of each party being $N=2^{L/2}$. Together with the original spin chain with spins indexed by $A_i$ and $B_j$ according to the notation explained before, we consider an ancillary spin chain of same length with sites labeled by  $A'_i$ and $B'_j$ ($i,j=1..L/2$), see Fig.~(\ref{fig:ancilla}).  We will be interested in the state,  
 \[ |\Phi \rangle_{ABA'B'}= \bigotimes_{i=1}^{L/2} |\Phi^+\rangle_{A_iA_i'} |\Phi^+\rangle_{B_iB_i'},\]
 where $|\Phi^+\kt$ is one of the Bell states:
\[ |\Phi^{\pm}\kt =\frac{1}{\sqrt{2}} ( |00\kt \pm |11\kt), \;\; |\Psi^{\pm}\kt= \frac{1}{\sqrt{2}} ( |01\kt \pm |10\kt).\] 
The linear $E_l(U^n \, S)$ is the linear entropy of the state, $((U^nS)_{AB} \otimes I_{A'B'}) |\Phi \rangle_{ABA'B'}$, across the $AA'|BB'$ partitions.
Notice that the swap operation has been clubbed along with the unitary evolution in this setting. Also it is sufficient here to consider only the nonlocal part $(U^n)_{nl}$ part of $U^n$.

We have,                 
\begin{equation}     
\left((U^n)_{nl}\, S\right)_{AB} \otimes I_{A'B'}) |\Phi \rangle_{ABA'B'}=\prod_{i=1}^nV_i \, \prod_{j=1}^{L/2}S_{jj}\, \bigotimes_{j=1}^{L/2}\,  |\Phi^+\rangle_{A_jA_j'} |\Phi^+\rangle_{B_jB_j'},
\end{equation}                 
where we have use that 
\[ S=\prod_{j=1}^{L/2}S_{jj}, \;\; \text{with}\;\; S_{ii} \equiv S_{A_iB_i}\] 
are swap operators on pairs of spins in the different partitions. Their action on the bell pairs yield
\begin{equation}
S_{ii}|\Phi^+\rangle_{A_iA_i'} |\Phi^+\rangle_{B_iB_i'} = |\Phi^+\rangle_{A_iB_i'} |\Phi^+\rangle_{A_i'B_i}  \equiv |\alpha\rangle_i,
\end{equation}
which has two ebits of entanglement (Schmidt rank $4$ state) across the $A_iA'_i|B_iB_i'$ partition. It is thus clear that, $|\alpha\rangle_1 \cdots |\alpha\rangle_{n}$ will be a 
maximally entangled state in $4^{n}$ ($2n$ ebits) dimensions. Particularly for $n=L/2$ we will have, $E_l(S)=1-1/2^L=1-1/N^2$.  
We need $E_l(U^nS)$ which is the linear entropy of the state, $\prod_{i=1}^n V_i \, \bigotimes_{j=1}^{L/2}|\alpha\rangle_j$.

Our basic strategy will consist of showing that the entanglement of the state,  $\prod_{i=1}^n V_i \bigotimes_{j=1}^n |\alpha\rangle_j$ grows as $n+1$ ebits for $n \leq L/2$.
On the other hand, $\bigotimes_{i=n+1}^{L/2} |\alpha\rangle_i$, will contribute $L-2n$ ebits, so that totally we have, $L-n+1$ ebits. 
This will imply, 
 $E_l(U^nS) = 1 - 2^{n-1}/2^L$. This together with the result for $E_l(U^n)$ proved before yields the desired result.
For, $n= L/2 + m$ ($m < L/2$) we will adopt the same strategy as for $E_l(U^n)$ for $n > L/2$  and define $V'$ operators. Then we would try to find the entanglement
in the state,
$\prod_{j=1}^{\frac{L}{2}-m} V_j \prod_{l=1}^{m} V'_l |\alpha\rangle_1|\alpha\rangle_2...|\alpha\rangle_{\frac{L}{2}}$. 
It will turn out that, this is in fact 
equal to the entanglement in the state, $\prod_{j=1}^{\frac{L}{2}-m} V_j  |\alpha\rangle_1|\alpha\rangle_2...|\alpha\rangle_{\frac{L}{2}-m}$ = $L/2-m+1 \mbox{ ebits }=L-n+1$ ebits. 

We state the following readily verifiable identities for later use:
\beq
\begin{split}
\label{eq:lemma1}
A_i^xB_i^x|\Phi^+\rangle_{A_iB_i'} |\Phi^+\rangle_{A_i'B_i} = |\Psi^+\rangle_{A_iB_i'} |\Psi^+\rangle_{A_i'B_i}\\
A_i^yB_i^y|\Phi^+\rangle_{A_iB_i'} |\Phi^+\rangle_{A_i'B_i} =  |\Psi^-\rangle_{A_iB_i'} |\Psi^-\rangle_{A_i'B_i} \\
A_i^zB_i^z|\Phi^+\rangle_{A_iB_i'} |\Phi^+\rangle_{A_i'B_i} =  |\Phi^-\rangle_{A_iB_i'} |\Phi^-\rangle_{A_i'B_i}. 
\end{split} 
\eeq
\beq
\begin{split}
\label{eq:lemma2}
|\Phi^+\rangle_{A_1B_1'} |\Phi^+\rangle_{A_1'B_1} +  |\Psi^-\rangle_{A_1B_1'} |\Psi^-\rangle_{A_1'B_1} = |\Phi^+\rangle_{A_1A_1'} |\Phi^+\rangle_{B_1B_1'} - |\Psi^-\rangle_{A_1A_1'} |\Psi^-\rangle_{B_1B_1'} \\
|\Phi^+\rangle_{A_1B_1'} |\Phi^+\rangle_{A_1'B_1} -  |\Psi^-\rangle_{A_1B_1'} |\Psi^-\rangle_{A_1'B_1} = |\Phi^-\rangle_{A_1A_1'} |\Phi^-\rangle_{B_1B_1'} + |\Psi^+\rangle_{A_1A_1'} |\Psi^+\rangle_{B_1B_1'} \\ 
|\Psi^+\rangle_{A_1B_1'} |\Psi^+\rangle_{A_1'B_1} +  |\Phi^-\rangle_{A_1B_1'} |\Phi^-\rangle_{A_1'B_1} = |\Phi^+\rangle_{A_1A_1'} |\Phi^+\rangle_{B_1B_1'} + |\Psi^-\rangle_{A_1A_1'} |\Psi^-\rangle_{B_1B_1'}  \\
|\Psi^+\rangle_{A_1B_1'} |\Psi^+\rangle_{A_1'B_1} -  |\Phi^-\rangle_{A_1B_1'} |\Phi^-\rangle_{A_1'B_1} = |\Psi^+\rangle_{A_1A_1'} |\Psi^+\rangle_{B_1B_1'}   -|\Phi^-\rangle_{A_1A_1'} |\Phi^-\rangle_{B_1B_1'}. 
\end{split}
\eeq

Let ${\tilde A_K}$ and ${\tilde B_K}$ represent following collections of spins: \[ {\tilde{A}}_K \equiv \{ A_1,A'_1, \cdots, A_K,\, A_K'\}, \;\;\; {\tilde{B}}_K \equiv \{ B_1, \, B'_1, \cdots  \, B_K, \, B_K'\}\] and  $C(n) = \prod_{k=1}^n A_k^xB_k^x$ be the string of operators.
We have 

{\textbf{Lemma 1:}} 
\begin{enumerate}
\item $V_1|\alpha\rangle_1$, has a Schmidt decomposition,
 \[ \frac{1}{2} \left( |00\rangle_{A_1A_1'}|\Phi'^{-}\rangle_{B_1B_1'} + i|01\rangle_{A_1A_1'} |\Psi'^{-}\rangle_{B_1B_1'} + 
|10\rangle_{A_1A_1'}|\Psi'^{+}\rangle_{B_1B_1'} - i|11\rangle_{A_1A_1'} |\Phi'^+\kt_{B_1B_1'} \right),\]
with $|\Phi'^{\pm}\rangle = \frac{1}{\sqrt{2}}(|00\rangle \pm i|11\rangle)$, and $|\Psi'^{\pm}\rangle = \frac{1}{\sqrt{2}}(|01\rangle \pm i|10\rangle)$. This can also be written in terms of bi-orthogonal vectors as 
\[ \frac{1}{2} \sum_{i=1}^2 \left(|e_i\rangle_{{\tilde{A}}_1{\tilde{B}}_1} + |e_i'\rangle_{{\tilde{A}}_1{\tilde{B}}_1} \right), \] so that $|e_i'\rangle _{\ta_1 \tb_1}=i \,C(1)|e_i\rangle_{\ta_1 \tb_1}$, where $|e_1\kt_{\ta_1 \tb_1} = |00\rangle_{A_1A_1'}|\Phi'^{-}\rangle_{B_1B_1'}$, and $|e_2\kt_{\ta_1 \tb_1} =- i|11\rangle_{A_1A_1'} |\Phi'^+\kt_{B_1B_1'}$.

\item $E_l(US)=E_l(S)$

\end{enumerate}
{\textbf{Proof:}} The first part follows from direct computations. We state the following useful identities:
\begin{eqnarray}
(\sigma_x \otimes I)|\Phi'^{\pm}\rangle = \pm i |\Psi'^{\mp} \rangle, \; (\sigma_x \otimes I)|\Psi'^{\pm}\rangle = \pm i |\Phi'^{\mp} \rangle .
\end{eqnarray}
and observe
\[ i|01\rangle_{A_1A_1'} |\Psi'^{-}\rangle_{B_1B_1'}\, =\, i \, C(1) \, (- i|11\rangle_{A_1A_1'} |\Phi'^{+}\rangle_{B_1B_1'}), \;\;
|10\rangle_{A_1A_1'}|\Psi'^{+}\rangle_{B_1B_1'}\, =\, i\, C(1)|00\rangle_{A_1A_1'}|\Phi'^{-}\rangle_{B_1B_1'}.\]
The second part follows as a consequence, as $V_1 |\alpha\kt_1$ is a rank-4 maximal Schmidt decomposition, the same as $|\alpha \kt_1$.  The von Neumann entanglement in these states is $2$ ebits. 

We are now ready to prove the main theorem.

\noindent{\textbf{Theorem 2 :}} 

The von Neumann entropy of $\prod_{i=1}^n V_i \bigotimes_{j=1}^n |\alpha\rangle_j$ is $=n+1$, $1 \le n\leq L/2$. 

\noindent{\textbf{Proof :}}

The proof is inductive. Let us first analyze, $V_2V_1|\alpha\rangle_1|\alpha\rangle_2 $. 
On expanding $V_2$ and using Eq.~(\ref{eq:lemma1}) we have,
\begin{equation}
 V_2V_1|\alpha\rangle_1|\alpha\rangle_2 = \frac{1}{\sqrt{2}}\left(V_1|\alpha\rangle_1|\Phi^+\rangle_{A_2B_2'} |\Phi^+\rangle_{A_2'B_2}- i\,C(1) V_1|\alpha\rangle_1|\Psi^-\rangle_{A_2B_2'}  
|\Psi^-\rangle_{A_2'B_2} \right).
\end{equation}
Let us now use the lemma above and  consider the contribution from the term $|e_i\rangle_{{\tilde{A}}_1{\tilde{B}}_1} + |e_i'\rangle_{{\tilde{A}}_1{\tilde{B}}_1}$, to the Schmidt decomposition of $V_2V_1|\alpha\rangle_1|\alpha\rangle_2$ for different $i$. We have on using $C(1)^2=I$, 
\begin{eqnarray}
V_2V_1|\alpha\rangle_1|\alpha\rangle_2 &=& \frac{1}{2\sqrt{2}} \sum_{i=1,2} \left[ |e_i\rangle_{{\tilde{A}}_1{\tilde{B}}_1} \left( |\Phi^+\rangle_{A_2B_2'} |\Phi^+\rangle_{A_2'B_2} +  |\Psi^-\rangle_{A_2B_2'} |\Psi^-\rangle_{A_2'B_2} \right)\right. \nonumber  \\
&+& \left.  |e_i'\rangle_{{\tilde{A}}_1{\tilde{B}}_1} \left(|\Phi^+\rangle_{A_2B_2'} |\Phi^+ \rangle_{A_2'B_2} -  |\Psi^-\rangle_{A_2B_2'} |\Psi^-\rangle_{A_2'B_2} \right)  \right] 
\end{eqnarray}
Using Eq.~(\ref{eq:lemma2}) we rewrite the state so that the partition $AA'|BB'$ or $\ta_2|\tb_2$ can be read off: 
\begin{eqnarray}
V_2V_1|\alpha\rangle_1|\alpha\rangle_2 &=& \frac{1}{2\sqrt{2}} \sum_{i=1,2} ( |e_i\rangle_{{\tilde{A}}_1{\tilde{B}}_1}
(|\Phi^+\rangle_{A_2A_2'} |\Phi^+\rangle_{B_2B_2'} - |\Psi^-\rangle_{A_2A_2'} |\Psi^-\rangle_{B_2B_2'} ) \nonumber  \\
&+&  |e_i'\rangle_{{\tilde{A}}_1{\tilde{B}}_1} (|\Phi^-\rangle_{A_2A_2'} |\Phi^-\rangle_{B_2B_2'} + |\Psi^+\rangle_{A_2A_2'} |\Psi^+\rangle_{B_2B_2'}).  
\end{eqnarray}
We started with a Schmidt decomposition of $V_1|\alpha\kt_1$, and clearly the above is also Schmidt decomposed across the $A_1A_2A_1'A_2'|  B_1B_2B_1'B_2'$ or $\ta_2|\tb_2$ partition. It is maximally entangled in a $8-$ dimensional subspace of the $16-$ dimensional space. Hence its von Neumann entropy is $3$ ebits. 

A crucial observation is that with suitable definitions this can be written as \[ V_2V_1|\alpha\rangle_1|\alpha\rangle_2=\frac{1}{2\sqrt{2}} \left(\sum_{i=1}^{4} 
|e_i\rangle_{{\tilde{A}}_2{\tilde{B}}_2} + |e_i'\rangle_{{\tilde{A}}_2{\tilde{B}}_2} \right),\] 
where for $i=1,2,$
\begin{eqnarray}
|e_i'\kt_{\ta_2 \tb_2}=|e_i'\rangle_{{\tilde{A}}_1{\tilde{B}}_1}|\Psi^+\rangle_{A_2A_2'}|\Psi^+\rangle_{B_2B_2'},& |e_i\kt_{\ta_2 \tb_2} &= |e_i\rangle_{{\tilde{A}}_1{\tilde{B}}_1}|\Phi^+\rangle_{A_2A_2'} |\Phi^+\rangle_{B_2B_2'} \\
|e_{i+2}'\kt_{\ta_2 \tb_2}=-|e_i\rangle_{{\tilde{A}}_1{\tilde{B}}_1}|\Psi^-\rangle_{A_2A_2'} |\Psi^-\rangle_{B_2B_2'}, & |e_{i+2}\kt_{\ta_2 \tb_2} &= |e_i'\rangle_{{\tilde{A}}_1{\tilde{B}}_1} |\Phi^-\rangle_{A_2A_2'} |\Phi^-\rangle_{B_2B_2'}. 
\end{eqnarray}
These satisfy $|e_i'\rangle_{{\tilde{A}}_2{\tilde{B}}_2}\, =\, i\, C(2)|e_i\rangle_{{\tilde{A}}_2{\tilde{B}}_2}$, for $1 \le i \le 4$ and this is exactly the relation between the Schmidt vectors at level $1$. Note that these continue to be bi-orthogonal and is therefore indeed a Schmidt decomposition.

\vspace{5mm}
Now, assume a Schmidt decomposition of rank $2^n$ of   
\begin{equation}
\left( \prod_{i=1}^{n-1} V_i \right) |\alpha\rangle_1|\alpha\rangle_2...|\alpha\rangle_{n-1}=
\frac{1}{2^{n/2}} \left(\sum_{i=1}^{2^{n-1}} (|e_i\rangle_{{\tilde{A}}_{n-1}{\tilde{B}}_{n-1}} + |e_i'\rangle_{{\tilde{A}}_{n-1}{\tilde{B}}_{n-1}})\right) \nonumber
\end{equation}
with 
$|e_i'\rangle_{{\tilde{A}}_{n-1}{\tilde{B}}_{n-1}}=iC(n-1)|e_i\rangle_{{\tilde{A}}_{n-1}{\tilde{B}}_{n-1}}$.  

Following the same steps as at the first level we get
\begin{eqnarray}
\prod_{i=1}^n V_i \bigotimes_{j=1}^n |\alpha\rangle_j
&=& \frac{1}{2^{(n+1)/2}} \sum_{i=1}^{2^{n-1}} ( |e_i\rangle_{{\tilde{A}}_{n-1}{\tilde{B}}_{n-1}}(|\Phi^+\rangle_{A_nA_n'} |\Phi^+\rangle_{B_nB_n'} - |\Psi^-\rangle_{A_nA_n'} |\Psi^-\rangle_{B_nB_n'} )
 \nonumber  \\
&+&  |e_i'\rangle_{{\tilde{A}}_{n-1}{\tilde{B}}_{n-1}}(|\Phi^-\rangle_{A_nA_n'} |\Phi^-\rangle_{B_nB_n'} + |\Psi^+\rangle_{A_nA_n'} |\Psi^+\rangle_{B_nB_n'})  ). 
\end{eqnarray}
which can also be written in Schmdit decomposed form using biorthogonal vectors as 
\begin{equation}
\label{eq:EUSleveln}
 \frac{1}{2^{(n+1)/2}}(\sum_{i=1}^{2^n} 
\left(|e_i\rangle_{{\tilde{A}}_n{\tilde{B}}_n} + |e_i'\rangle_{{\tilde{A}}_n{\tilde{B}}_n} \right), 
\end{equation}

where for $1 \le i \le 2^{n-1}$
\begin{eqnarray}
|e_i'\kt_{\ta_n \tb_n}=|e_i'\rangle_{{\tilde{A}}_{n-1}{\tilde{B}}_{n-1}}|\Psi^+\rangle_{A_nA_n'}|\Psi^+\rangle_{B_nB_n'},& |e_i\kt_{\ta_n \tb_n} &= |e_i\rangle_{{\tilde{A}}_{n-1}{\tilde{B}}_{n-1}}|\Phi^+\rangle_{A_nA_n'} |\Phi^+\rangle_{B_nB_n'} \\
|e_{i+2^{n-1}}'\kt_{\ta_n \tb_n}=-|e_i\rangle_{{\tilde{A}}_{n-1}{\tilde{B}}_{n-1}}|\Psi^-\rangle_{A_n A_n'} |\Psi^-\rangle_{B_n B_n'}, & |e_{i+2^{n-1}}\kt_{\ta_n \tb_n} &= |e_i'\rangle_{{\tilde{A}}_{n-1}{\tilde{B}}_{n-1}} |\Phi^-\rangle_{A_n A_n'} |\Phi^-\rangle_{B_n B_n'}. 
\end{eqnarray}
These satisfy $|e_i'\rangle_{{\tilde{A}}_n{\tilde{B}}_n}\, =\, i\, C(n)|e_i\rangle_{{\tilde{A}}_n{\tilde{B}}_n}$, for $1 \le i \le 2^n$, the same relation at the previous levels.
Thus it follows that the von Neumann entropy of $ \prod_{i=1}^{n} V_i |\alpha\rangle_1|\alpha\rangle_2...|\alpha\rangle_{n}$ is $n+1$.

\vspace{5mm}
{\textbf{Theorem 3 :}} 
 The von Neumann entropy of the state $ \prod_{j=1}^{\frac{L}{2}+m} V_j|\alpha\rangle_1|\alpha\rangle_2...|\alpha\rangle_{\frac{L}{2}})$ is equal to the von Neumann entropy of the state $ |\beta_m\kt \equiv \prod_{j=1}^{\frac{L}{2}-m} V_j  |\alpha\rangle_1|\alpha\rangle_2...|\alpha\rangle_{\frac{L}{2}-m}$, \, for $1 \leq m < L /2$.

\vspace{5mm}
{\textbf{Proof :}}

We use again the strategy of pairing the $V_j$ operators beyond $j=L/2-m$ into $V'_{m+1-k}=V_{L/2-k+1} V_{L/2+k}$. Denote $n_2=n_1+1=L/2-m+1$, and in all expressions $1 \leq m < L/2$. For example, 
\begin{equation}
V'_{1}=V_{\frac{L}{2}-m+1}V_{\frac{L}{2}+m} = \frac{1}{2}(I-iA^z_{n_2} B^z_{n_2}C(n_1) -i A^y_{n_2} B^y_{n_2}C(n_1) +  A^x_{n_2}   B^x_{n_2}) .  
\end{equation}
Direct computation and the usage of identities in Eq.~(\ref{eq:lemma2}) results in  
\[ V'_{1}\, |\beta_m \rangle |\alpha\rangle_{n_2}= \frac{1}{2^{(n_1+1)/2}}\sum_{i=1}^{2^{n_1}} \left(
|e_i\rangle_{{\tilde{A}}_{n_1}{\tilde{B}}_{n_1}}|\Phi^+\rangle_{A_{{n_2}}B_{n_2}'} |\Phi^+\rangle_{A_{n_2}'B_{n_2}} +
|e_i'\rangle_{{\tilde{A}}_{n_1}{\tilde{B}}_{n_1}}|\Psi^+\rangle_{A_{{n_2}}B_{n_2}'} |\Psi^+\rangle_{A_{n_2}'B_{n_2}}\right).\]
As previously we have, 
\[ |e_i'\rangle_{{\tilde{A}}_{n_1}{\tilde{B}}_{n_1}}|\Psi^+\rangle_{A_{{n_2}}B_{n_2}'} |\Psi^+\rangle_{A_{n_2}'B_{n_2}}=
iC(n_2)|e_i\rangle_{{\tilde{A}}_{n_1}{\tilde{B}}_{n_1}}|\Phi^+\rangle_{A_{{n_2}}B_{n_2}'} |\Phi^+\rangle_{A_{n_2}'B_{n_2}}.\]  
But note that $|\beta_m\kt$ is also in Eq.~(\ref{eq:EUSleveln}) with $n=n_1 =L/2-m$.
Hence,  $V'_{1}|\beta_m\rangle |\alpha\rangle_{n_1+1}$ has the same entanglement as $|\beta_m\rangle$.
Now, assuming $\prod_{l=1}^k V'_l |\beta_m\rangle |\alpha\rangle_{n_1+1}|\alpha\rangle_{n_1+2}..|\alpha\rangle_{n_1+k}$ has the same 
entanglement  as $|\beta_m\rangle$ ($k\leq m$) it is easy to show by following the same steps that, $\prod_{l=1}^{k+1} V'_l |\beta_m\rangle |\alpha\rangle_{n_1+1}|\alpha\rangle_{n_1+2}..|\alpha\rangle_{n_1+k+1}$ also has the same 
entanglement. Hence, the theorem follows by induction. 

\noindent {\textbf{Theorem 4 :}} 
For $1 \le n \le L$, the linear operator entanglement entropy is $E_l(U^nS) = 1 - 2^{-L+n-1}$,
 
\noindent{\textbf{Proof :}} 
Let us denote the von Neumann entropy of a state $|\cdot \kt$, by $\mathcal{E}_{vn}(|\cdot \kt)$. For $n \leq L/2$, by Theorem~2,
 \[\evn \left(\prod_{i=1}^n V_i \bigotimes_{j=1}^n |\alpha\rangle_j\right)=n+1.\] 
We also have,
\[\evn\left(\bigotimes_{i=n+1}^{L/2} |\alpha\rangle_i\right)=2 \, (L/2-n)=L-2n.\] 
Hence,
\begin{equation}
\evn \left(\prod_{i=1}^n V_i \bigotimes_{j=1}^n |\alpha\rangle_j\bigotimes_{i=n+1}^{L/2} |\alpha\rangle_i \right) =(L-2n)+ (n+1)=L-n+1 \nonumber 
\end{equation}
As the Schmdit decompositions are all maximal, the linear operator entanglement is
\begin{equation}
\label{eq:eus}
E_l \left(U^nS \right)=1-\frac{1}{2^{(L-n+1)}}=1 - \frac{ 2^{n-1}}{ 2^L}.
\end{equation}

For times larger than $n=L/2$, consider $L/2 < n < L$, and let $m=n-L/2$.
It follows from Theorem~3, that 
\begin{equation}
\label{eq:vNUnS}
\evn \left(\prod_{j=1}^{\frac{L}{2}+m} V_j|\alpha\rangle_1|\alpha\rangle_2...|\alpha\rangle_{\frac{L}{2}}\right) = L/2-m+1 =L-n+1. 
\end{equation}
Hence Eq. (\ref{eq:eus}) continues to hold. We still have one more time $n=L$ to cover and this needs special treatment. 
We need to find 
\[ \evn\left (\prod_{l=1}^{L/2} V'_l \bigotimes_{j=1}^{L/2} |\alpha \kt_j  \right).\] 

By direct, but tedious, computation it follows that,
\[ V'_1|\alpha\kt_1= \frac{1}{\sqrt{2}}\left( \frac{(1-i)}{\sqrt{2}}|\Phi^+\rangle_{A_1A_1'} |\Phi^+\rangle_{B_1B_1'} + \frac{(1+i)}{\sqrt{2}} |\Psi^+\rangle_{A_1A_1'} |\Psi^+\rangle_{B_1B_1'}  \right ) \] and
\[\frac{(1+i)}{\sqrt{2}} |\Psi^+\rangle_{A_1A_1'} |\Psi^+\rangle_{B_1B_1'} = iA_1^xB_1^x  \frac{(1-i)}{\sqrt{2}} |\Phi^+\rangle_{A_1A_1'} |\Phi^+\rangle_{B_1B_1'}.  \]
Hence, following the same steps as in the proof of Theorem~3, it is straightforward to see that
\[ \evn \left(\prod_{l=1}^{L/2} V'_l \bigotimes_{j=1}^{L/2} |\alpha  \kt_j \right)= \evn\left( V'_1|\alpha\kt_1\right)= 1.  \]
Thus Eq.~(\ref{eq:vNUnS}) holds uniformly for $1 \le n \le L$. In particular note that it decreases from $L$ initially ($n=0$) and at $n=1$ to $1$ at $n=L$. This precipitous fall makes up for the rise of the operator entanglement of $U^n$.
This also completes the proof of the theorem. 

For $n \geq L$, we again use the fact that $U^{2L}$ is a local operator. We have, $E_l(U^{2L-m}S)=E_l(U^{-m}S)=E_l(U^{m}S)$. Now, putting $m=L-m$, we have $E_l(U^{L-m}S)=E_l(U^{L+m}S)$.

\noindent{\textbf{Theorem 5 :}} The linear entropy entangling power is 
$\ep_l(U^n)=\dfrac{1+2^L -2^{L-n}-2^{n-1}}{(1+2^{L/2})^2}$, $ 1\le n \le L$, with $\ep_l(U^0)=0$ \\
{\textbf{Proof :}} 
This follows after elementary manipulations, using Theorems~1~and~4 together with Eq.~(\ref{epu}) and $E_l(S)=1-1/2^L$.

\section{Anti-unitary symmetry and false time reversal}

Suppose, the system has an anti-unitary symmetry governed by $T=G\, K$, where $K$ is the complex conjugation operator and $G$ is unitary. The condition that time-reversal like symmetry holds for a system whose time evolution is the unitary operator $U$, is \cite{Haake}
\[ T\, U\, T^{-1}\, =\, U^{-1}\, =\, U^{\dagger},\] implying that
\begin{equation}
\label{trs}
GU^*G^{-1}= U^{\dagger}.  
\end{equation}

For our model we have, 
\begin{equation}
U^*=\exp{\left(i\tau \left(\sum_{j=1}^{L-1} \sigma_j^z \sigma_{j+1}^z + \sum_{j=1}^L h_i^z \sigma_i^z\right)\right)} \exp{\left(i\tau\sum_{j=1}^L \left(h_j^x\sigma_j^x -  h_j^y \sigma_j^y \right) \right)} \nonumber
\end{equation}
and 
\begin{equation}
U^{\dagger}= \exp{\left(i\tau\sum_{j=1}^L \left(h_j^x\sigma_j^x +  h_j^y \sigma_j^y \right) \right)} \exp{\left(i\tau \left(\sum_{j=1}^{L-1} \sigma_j^z \sigma_{j+1}^z + \sum_{j=1}^L h_i^z \sigma_i^z\right)\right)} .
\end{equation}

For, $h_j^y=0$ for all $j$ clearly, 
\begin{equation}
G_1=\exp{\left(-i\tau \left( \sum_{j=1}^{L-1} \sigma_j^z \sigma_{j+1}^z + \sum_{j=1}^L h_i^z \sigma_i^z\right)\right)},
\end{equation} satisfies Eq.~(\ref{trs}).

We have,
\begin{equation}
G_1U^*G_1^{-1}=   \exp{\left(i\tau\sum_{j=1}^L \left(h_j^x\sigma_j^x -  h_j^y \sigma_j^y \right) \right)} \exp{\left(i\tau \left(\sum_{j=1}^{L-1} \sigma_j^z \sigma_{j+1}^z + \sum_{j=1}^L h_i^z \sigma_i^z\right)\right)}
\end{equation}

Let $h_j^x \, \hat{x} - h_j^y \, \hat{y}= h_j (\cos(\theta_j)\hat{x} - \sin(\theta_j)\hat{y})=h_j \hat{h_j} $ and $V_j=\exp{(-i\sigma_j^z\theta_j)}$ be a spin rotation operator, performing the rotation $2\theta_j$ about z- axis so that we have,
 
\begin{equation}
V_j(\cos(\theta_j)\sigma_j^x -  \sin(\theta_j) \sigma_j^y)V_j^{\dagger}=(\cos(\theta_j)\sigma_j^x +  \sin(\theta_j) \sigma_j^y) , \hspace{3mm}  [V_j,\sigma_j^z]=0. \nonumber 
\end{equation}

It follows that,
\begin{equation}
V_j \exp{\left(i\tau \left(h_j^x\sigma_j^x -  h_j^y \sigma_j^y \right)\right)}V_j^{\dagger}= \exp{ \left(i\tau \left( h_j^x\sigma_j^x +  h_j^y \sigma_j^y \right) \right)}.   
\end{equation}

Hence, with \[ G=\left(\bigotimes_{j=1}^{L} V_j\right) \,  G_1\]  Eq.~(\ref{trs}) is satisfied for the Floquet operator. This implies COE statistics as we have shown, and is valid for all disordered models as well.

%
%

\end{document}